\documentclass[conference]{IEEEtran}
\ifCLASSINFOpdf
\else
\fi
\usepackage[font=footnotesize]{subfig}
\usepackage{array}
\usepackage[cmex10]{amsmath}
\usepackage{cite}
\usepackage{multicol}
\usepackage{fixltx2e}
\usepackage{stfloats}
\usepackage{url}
\usepackage{tikz}
\usepackage{bbm}
\usepackage{subfig}
\usepackage[indLines=false]{algpseudocodex}
\usepackage{algorithm}
\usepackage{xcolor}
\usepackage{colortbl}  
\usepackage{xspace}
\usepackage{balance}
\usepackage{nicefrac}
\usepackage{adjustbox}
\usepackage{amsmath}
\usepackage{booktabs}
\usepackage[nameinlink,capitalise]{cleveref}
\PassOptionsToPackage{table}{xcolor}
\usepackage[numbers,sort&compress]{natbib}

\algblockdefx[game]{game}{EndGame}[2]{\textbf{experiment} \textsc{#1}(#2)}{\algpx@endIndent}
\algtext*{EndGame}
\usepackage{wrapfig}
\usepackage{multirow}
\let\oldReturn\Return
\renewcommand{\Return}{\State\oldReturn}
\usepackage{tcolorbox}

\definecolor{redbg}{RGB}{254,241,240}
\definecolor{redoutline}{RGB}{252,163,152}
\definecolor{redtext}{RGB}{207,24,34}
\definecolor{beigeoutline}{RGB}{220, 220, 200} %
\definecolor{beigebg}{RGB}{245, 245, 220} %
\definecolor{greybg}{RGB}{220, 220, 230} %
\DeclareRobustCommand*{\cleanlabel}[1]{%
  \tikz[baseline=(char.base)]\node[anchor=south west, draw, rectangle, thick, rounded corners=0.2mm, inner sep=2pt, fill=beigebg, draw=beigeoutline,text=black](char){#1} ;}

  \DeclareRobustCommand*{\poisonlabel}[1]{%
  \tikz[baseline=(char.base)]\node[anchor=south west, draw, rectangle, thick, rounded corners=0.2mm, inner sep=2pt, fill=redbg, draw=redoutline,text=black](char){#1} ;}

\DeclareRobustCommand*{\supplychain}[1]{%
  \tikz[baseline=(char.base)]\node[anchor=south west, draw, rectangle, thick, rounded corners=0.2mm, inner sep=2pt, fill=greybg, draw=gray,text=black](char){#1} ;}

\usepackage{amssymb}

\begin{document}
\title{Pick your Poison: Undetectability versus Robustness in Data Poisoning Attacks}

% use a multiple column layout for up to three different% affiliations
\author{\IEEEauthorblockN{Nils Lukas}
\IEEEauthorblockA{University of Waterloo\\
nlukas@uwaterloo.ca}
\and
\IEEEauthorblockN{Florian Kerschbaum}
\IEEEauthorblockA{University of Waterloo\\
florian.kerschbaum@uwaterloo.ca}
}

%\author{\IEEEauthorblockN{Anonymous Author(s)*}
%\vspace*{0ex}
%\IEEEauthorblockA{ x \\ x}
%}


%Telephone: (800) 555--1212, Fax: (888) 555--1212}


\IEEEoverridecommandlockouts
\makeatletter\def\@IEEEpubidpullup{6.5\baselineskip}\makeatother
\IEEEpubid{\parbox{\columnwidth}{
    This paper is made available as a preprint.\\
    ~
    \\~
    \\~
    \\~
}
\hspace{\columnsep}\makebox[\columnwidth]{}}

\maketitle

\begin{abstract}
Deep image classification models trained on vast amounts of web-scraped data are susceptible to data poisoning—a mechanism for backdooring models.
A small number of poisoned samples seen during training can severely undermine a model's integrity during inference.
Existing work considers an effective defense as one that either (i) restores a model's integrity through repair or (ii) detects an attack.
We argue that this approach overlooks a crucial trade-off: attackers can increase robustness at the expense of detectability (over-poisoning) or decrease detectability at the cost of robustness (under-poisoning).
In practice, attacks should remain both undetectable \emph{and} robust.
Detectable but robust attacks draw human attention and rigorous model evaluation or cause the model to be re-trained or discarded. 
In contrast, attacks that are undetectable but lack robustness can be repaired with minimal impact on model accuracy. 
Our research points to intrinsic flaws in current attack evaluation methods and raises the bar for all data poisoning attackers who must delicately balance this trade-off to remain robust and undetectable.
To demonstrate the existence of more potent defenders, we propose defenses designed to (i) detect or (ii) repair poisoned models using a limited amount of trusted image-label pairs. 
Our results show that an attacker who needs to be robust and undetectable is substantially less threatening.
Our defenses mitigate all tested attacks with a maximum accuracy decline of $2\%$ using only $1\%$ of clean data on CIFAR-10 and $2.5\%$ on ImageNet. 
We demonstrate the scalability of our defenses by evaluating large vision-language models, such as CLIP. 
Although our results indicate that data poisoning might be less threatening than previously believed, we recognize that backdooring remains a significant threat. 
Attackers who can manipulate the model's parameters pose an elevated risk as they can achieve higher robustness at low detectability compared to data poisoning attackers.
\end{abstract}

\section{Introduction}
The integrity of deep image classification models is essential for their safe and effective use in real-world applications~\cite{papernot2018sok}.
A model with integrity should make accurate, transparent, and robust decisions that are resistant to manipulation.
Training models with integrity is challenging, which limits the broader applicability of models in domains with impactful decision-making like diagnostic healthcare~\cite{biggio2012biometricpoisoning, rasheed2022explainable}, credit scoring~\cite{bucker2022transparency}, and autonomous driving~\cite{shen2022sok}.
% autonomous driving: shen2022sok

One form of manipulation that can undermine a model's integrity is \emph{backdooring} - an attack during training enabling the attacker to influence the model's predictions during inference by stamping a secret \emph{trigger} pattern on an image~\cite{chen2017targeted}.
Backdooring can be accomplished through \emph{data poisoning}, a technique in which the attacker subtly injects a limited number of poisoned samples into the model's training data, thereby creating spurious correlations~\cite{Sagawa2020Distributionally} between the trigger and an attacker-chosen \emph{target} class. 
Given that modern deep learning models rely on vast amounts of web-scraped data for optimal accuracy~\cite{radford2021learning}, poisoning attacks are considered among the most worrisome threats to a model's integrity in practice~\cite{kumar2020adversarial}.

\textbf{Problem.} Consider a provider who trains a deep image classifier using web-scraped data and deploys their model for a downstream task such as content moderation, like OpenAI with CLIP~\cite{radford2021learning}.
The provider, being trustworthy, must balance the trade-off between high model accuracy and integrity but needs to train on large, web-scraped datasets to achieve the former.
The threat is an attacker who poisons a small number of samples during training to evade content moderation during inference. 
With CLIP, it has been demonstrated that backdooring succeeds by poisoning only $400$ out of $400$ million image-label pairs~\cite{radford2021learning}. 
The provider needs a method to mitigate data poisoning while training an accurate model. 

To counter backdooring, the provider can apply defenses before or after training. 
Before training, the provider can (i) filter how data is collected~\cite{carlini2023poisoning}, (ii) sanitize data using trusted curators~\cite{borgnia2021dp, Tran2018SpectralSI, chen2018detecting, shan2022poison} or (iii) use algorithmic defenses to train robust models~\cite{du2019robust, liu2018fine, li2021anti, huang2022backdoor, dolatabadi2022collider}. 
After training, providers can evaluate their models to (v) detect unwanted behavior~\cite{wang2019neural,guo2019tabor} or (vi) pre-emptively repair models suspected of containing a backdoor~\cite{liu2018fine,li2021neural} using limited high-quality data from trusted sources.
However, none of these methods have been shown to withstand existing attacks~\cite{wu2022backdoorbench,doan2021backdoor}. 
Data sanitation defenses have been broken by stronger attacks~\cite{doan2021lira,koh2022stronger}, algorithmic defenses have only been tested on small-scale datasets with less than $100$k samples, and post-training defenses either fail to detect attacks or fail to repair the model while preserving accuracy~\cite{wu2022backdoorbench}. 
The vulnerability to data poisoning has called into question whether web-scraped training is at all desirable~\cite{carlini2022poisoning}. 

\textbf{Overview.} 
An attack's undetectability and its robustness are studied in isolation~\cite{wang2019neural,li2021neural,liu2018fine, wu2022backdoorbench}, which we argue underestimates the  effectiveness of defenses.
We make the observation that an attack's undetectability and robustness are at odds: attackers who \emph{over-poison} by injecting too many samples increase robustness at the expense of higher detectability, and vice-versa, attackers who \emph{under-poison} reduce detectability at the cost of lower robustness.
In practice, an attacker wants to remain \emph{both} undetectable and robust and needs to optimize their attack to achieve both goals.
A robust but easily detectable attack is undesirable because it triggers rigorous model evaluation by humans, increases scrutiny of the model's outputs for the target class, or causes the model to be rejected or re-trained~\cite{gpt4tech}. 
Similarly, an undetectable attack that lacks robustness can easily be removed with limited impact on model accuracy through pre-emptive model repair.  

To demonstrate that more potent defenders can be instantiated, we develop improved defenses for (i) backdoor detection and (ii) model repair that require less trusted data than existing methods while better preserving the model's accuracy. 
The idea of our model repair defense is to use a regularization technique called Pivotal Tuning~\cite{roich2022pivotal} to maximize the model's dissimilarity to a frozen clone of itself while preserving its accuracy on a small set of trusted data. 
We measure dissimilarity on the model's \emph{latent space}, which is a lower-dimensional representation of images learned during training that is often used to interpret the model~\cite{ramesh2022hierarchical}. 
While there are methods to enhance latent representations through regularization~\cite{frosst2019analyzing} or latent space alignment~\cite{morcos2018insights}, no methods exist to maximize the \emph{dissimilarity} between two latent spaces to repair backdoors. 

The idea behind our backdoor detection method is to measure whether a model suspected of containing a backdoor is more sensitive to changes in the input than a repaired version of this model. 
We optimize to reconstruct the secret trigger patterns through iterative optimization and output anomaly scores for each class. 
Existing detection methods, such as Neural Cleanse~\cite{wang2019neural}, easily get stuck in local minima, which leads to a high false negative detection rate. 
This issue is caused by the model's vulnerability to universal adversarial examples~\cite{moosavi2017universal}, where a random adversarial trigger has similar properties as a maliciously inserted trigger, which makes detection of the latter more difficult.
Our optimization bypasses this issue because both the backdoored and repaired models remain vulnerable to similar adversarial examples but differ in their behavior on maliciously inserted triggers.  

\textbf{Results.} Our empirical evaluation shows that our defenses can mitigate data poisoning attacks that neither over- nor under-poison. 
We show that our defense can repair all surveyed backdoored models on CIFAR-10~\cite{cifar10} using only 1\% of trusted data and 2.5\% on ImageNet~\cite{imagenet} with a maximum degradation in test accuracy of $2\%$,  outperforming all other surveyed post-training defenses. 
By ablating over the number of poisoned samples, we observe a fundamental trade-off between robustness and undetectability that allows crafting such effective defenses.
Interestingly, we find that detection-based defenses such as Neural Cleanse~\cite{wang2019neural} and our detection method are successful at detecting backdoors in larger models but often fail to reconstruct the \emph{exact} trigger needed to repair the model. 
In contrast, on smaller models, Neural Cleanse often fails to detect an attack but can reconstruct the trigger with high fidelity when it is successful (which is useful for repairing the model). 
We expand our evaluation to state-of-the-art joint vision-language models~\cite{radford2021learning}, demonstrating our post-training defenses' effectiveness and data efficiency. 

Finally, we show that more capable attackers who control the model's parameters can perform attacks with higher robustness and lower detectability than a data poisoning attacker. 
While it has been shown that such attackers can plant provably undetectable backdoors~\cite{goldwasser2022planting}, the robustness of undetectable backdoors is unknown. 
We propose an attack called the \emph{Parameter-Controlled Backdoor} (PCB), in which an attacker inserts malicious backdoor functionality only into a small subset of the model's parameters, allowing them to achieve low detectability at high robustness.  

% --------------------------------------------------------------------
\subsection{Contributions}
% --------------------------------------------------------------------
Our key contributions can be summarized as follows.
\begin{enumerate} \itemsep0mm
    \item We demonstrate that undetectability and robustness are at odds, raising the bar for all data poisoning attacks that must balance this trade-off. 
    \item We propose rigorous, game-based definitions of robustness and detectability for data poisoning. 
    \item We propose defenses to (i) repair and (ii) detect backdoored models that outperform existing defenses by achieving a better accuracy/integrity trade-off. 
    \item 
    Our defenses require less trusted data to repair backdoored models on CIFAR-10 and ImageNet. Our methods use only 1\% and 2.5\% of the training data with a deterioration of less than 2\% test accuracy. 
    \item We show results on large pre-trained models such as OpenAI's CLIP classifier~\cite{radford2021learning}. Our results indicate that attacks on large models and datasets are easier to detect but also more robust. 
    \item We show that a more capable attacker who manipulates the model's parameters can perform substantially more threatening backdoor attacks.  
    \item We release open source code\footnote{Blinded during review} to reproduce all our experiments, including all hyperparameters. 
\end{enumerate}

% ==========================================================
\section{Background }
% ==========================================================
\begin{figure*}
    \centering
    \includegraphics[width=1.\linewidth]{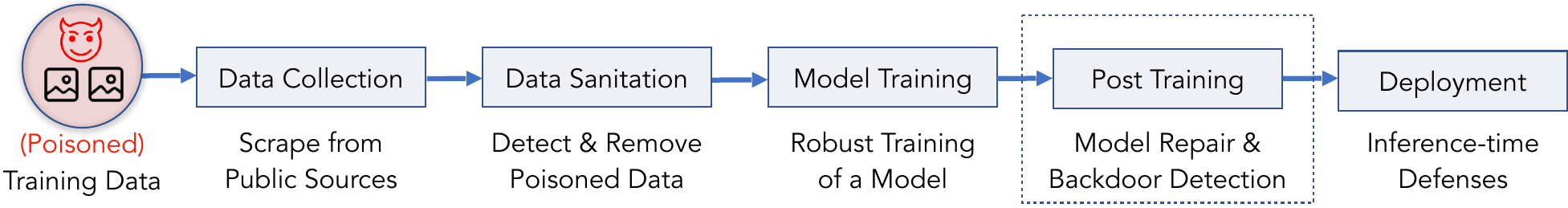}
    \caption{An exemplary training pipeline for deep image classification models from data collection to deployment. We focus on detection and model repair methods that are applied post-training.}
    \label{fig:training_pipeline}
\end{figure*}
This section provides a comprehensive overview of backdooring deep image classifiers. We define a dataset as $D \subseteq \mathcal{X}$, with $\mathcal{X}$ representing the set of images. A labeling function $\mathcal{O}: \mathcal{X} \rightarrow \mathcal{Y}$ maps images to labels within the set $\mathcal{Y}$. The training process aims to build a model that accurately replicates the labeling function for any image dataset. A \emph{poisoned} image refers to data that has been maliciously manipulated to cause a misclassification during inference.
% ----------------------------
\subsection{Data Poisoning Attacks}
\label{sec:data_poisoning_attacks}
% ----------------------------
Data poisoning attacks involve manipulating a small subset of the training data with the objective of compromising a model's integrity during inference. 
Here, we focus on \emph{targeted} backdoor attacks, where the adversary's primary goal is to remain undetected while being able to induce a specific misclassification during test time by modifying the input. 
These attacks can be broadly categorized into poison-label and clean-label attacks.

Poison-label attacks grant the attacker limited control over images and the labeling function~\cite{gu2017badnets}, whereas clean-label attacks assume that an attacker only has the capability to manipulate the images~\cite{shafahi2018poison}.
A central concept in these attacks is that of a \emph{trigger}, which is an attacker-chosen pattern introduced into the data that produces the desired misclassification when recognized by the model. 
Triggers are typically unique to the attack and are designed to be as inconspicuous as possible.

\poisonlabel{Poison-Label}. In Badnets~\cite{gu2017badnets}, the attacker uses a visible trigger patch during training and assigns the target class label to all poisoned images. 
More subtle techniques like Adaptive Blend (A-Blend) and Adaptive Patch (A-Patch)~\cite{qi2023revisiting} assign the target label only to a subset of all poisoned images.

\cleanlabel{Clean-Label}\footnote{We use colors to indicate clean-label and poison-label attacks.}. Clean-label attacks such as those proposed by Shafahi et al.~\cite{shafahi2018poison} use an adversarially crafted trigger imprinted on poisoned images, assuming access to the attacked model's parameters. 
Other attacks, such as the one presented by Turner et al.~\cite{turner2018clean}, do not require knowledge of the attacked model's parameters and apply adversarial triggers generated using a different model trained for the same task. 
Techniques like Refool~\cite{liu2020reflection} exploit naturally occurring trigger patterns, like reflections, while WaNet~\cite{nguyen2021wanet} uses unnoticeable warping of the image. 
These clean-label attacks utilize the same underlying principle of poisoning samples from the target class, but the trigger generation process varies across them.

We survey six data poisoning attacks from related work~\cite{turner2018clean,nguyen2021wanet,qi2023revisiting,liu2020reflection,gu2017badnets} (Qi et al.~\cite{qi2023revisiting} propose two backdoors).
Detailed descriptions of all surveyed attacks, their hyper-parameters, and illustrations of the triggers used are moved to \Cref{appendix:surveyed_attack_summary} for brevity.

% ----------------------------------------------------
\subsection{Data Poisoning Defenses}
\label{sec:data_poisoning_defenses}
% ----------------------------------------------------
\Cref{fig:training_pipeline} demonstrates the training pipeline for a deep image classification model using web-scraped data. 
The training data is collected from public sources and could include a small proportion of poisoned images~\cite{carlini2023poisoning}. 
Dataset sanitation processes attempt to filter or augment these poisoned images~\cite{chen2018detecting, hayase2021spectre, schulth2022detecting}, followed by a robust training approach~\cite{wangtraining,borgnia2021dp,huang2022backdoor,li2021anti}. 
After training, if backdooring is suspected, the defender can opt to inspect or repair the model using a limited amount of trustworthy data~\cite{wang2019neural, li2021nad, liu2018fine} before deployment. 
The final line of protection includes measures implemented during inference, such as anomaly detection systems~\cite{udeshi2022model}, input purification~\cite{doan2020februus, may2023salient} or certifiable defenses~\cite{steinhardt2017certified, zhang2022bagflip, jia2022certified, xiang2022patchcleanser}.

\textbf{Post-Training Defenses.}
We highlight four post-training defenses for deep image classification models with high effectiveness~\cite{wu2022backdoorbench}.
Weight Decay~\cite{loshchilov2017decoupled}, a well-known regularization method, penalizes the model for learning weights with a large magnitude.
Fine-Pruning~\cite{liu2018fine} prunes convolutional filters on a channel-wise basis if their absolute activation on clean data is low. 
Neural Attention Distillation~\cite{li2021neural} (NAD) fine-tunes a teacher model using the poisoned model as an initial point and distills its activations to a student model. 
 Finally, Neural Cleanse~\cite{wang2019neural} reconstructs the trigger pattern by backpropagating through the model and optimizing for a mask that behaves like a malicious trigger. 
 The backdoor can be repaired by fine-tuning with masked images and ground-truth labels. 
% ============================
\section{Threat Model}
\label{sec:threat-model}
% ===========================
This section describes our threat model and introduces rigorous security games for an attack's robustness and undetectability properties. 
To this end, we introduce the following functions corresponding to the training pipeline of a deep classification model depicted in \Cref{fig:training_pipeline}.
\begin{itemize}
    \itemsep0mm
    \item \textsc{Sanitize}($\hat{D}, \hat{Y}$): Returns a filtered and potentially augmented set of images $D$ and labels $Y$ given a set of uncurated images $\hat{D}$ and labels $\hat{Y}$. 
    \item \textsc{Train}($D, Y; \theta_{\text{init}}$): A stochastic training algorithm that learns a model from image-label pairs. This function optionally accepts pre-trained weights $\theta_{\text{init}}$.
    \item \textsc{Repair}($\theta, D, Y$): Given a pre-trained model $\theta$ and labelled samples, this method returns modified model parameters $\hat{\theta}$ with the aim to repair backdoors.
    \item \textsc{Deploy}($\theta$): Given a model $\theta$, this method modifies the inference function of the model.
    \item \textsc{Detect}($\theta, D, Y$): Outputs $0$ for backdoored models and $1$ for clean models, given a labelled, non-poisoned dataset $D, Y$ and model $\theta$.
    \item $\mathcal{A}(D, \mathcal{O})$: Given images $D$ and a ground-truth labeling function $\mathcal{O}$, this function returns poisoned images and the attack's target labels for each poisoned image. 
    \item \textsc{Acc}($D, Y;\theta$): Given a set of image-label pairs $(D,Y)$ and a labeling function on $D$ parameterized by $\theta$, this function computes the model's accuracy against $Y$. 
\end{itemize}
\Cref{tab:notation} summarizes our notation.

\begin{table}
    \caption{A summary of our notation.}
    \label{tab:notation}
    \centering
    \small
    \begin{tabular}{@{}p{63pt}@{~~}l@{}}
    \toprule
    \bf Notation                         & \bf Description \\
    \midrule
    $\theta$                             & Parameters of a deep image classification model \\
    $\mathcal{D}$                        & A distribution over images\\
    $D \sim \mathcal{D}^n$               & Draw $n$ independent images $D$ from $\mathcal{D}$ \\
    $\mathcal{O}$                        & A ground-truth labelling function for images \\
    $\mathcal{A}$                        & A procedure denoting an adversary \\
    $\nicefrac{m}{n}$                    & Ratio between poisoned and clean samples \\
    $r$                                  & The number of trusted, clean image-label pairs \\
    $\eta$                              & The Attack Success Rate (ASR) \\
    $y \gets \mathcal{P}(\vec{x})$       & Call $\mathcal{P}$ with arguments $\vec{x}$ and assign result to $y$ \\
    \bottomrule
    \end{tabular}
\end{table}

\textbf{Adversary's Capabilities and Goals.} 
We consider an attacker capable of poisoning a limited number of $m$ samples in the defender's training dataset. 
The attacker's goal is to provoke a misclassification of a sample during inference into a pre-defined \emph{target} class by stamping the secret trigger pattern on the image. 
The adversary succeeds when the model incorrectly predicts the target class, diverging from the ground truth label (reflecting the backdoor functionality) while concurrently maintaining accurate predictions for non-manipulated, clean images (embodying the benign functionality).
In such a targeted attack, the adversary has the incentive to preserve the model's accuracy to (i) remain stealthy and (ii) maximize the chance that the defender deploys this backdoored classifier.  

\textbf{Defender's Capabilities and Goals.} 
Our defender controls the entire training pipeline of their model from data collection to deployment as depicted in \Cref{fig:training_pipeline}. 
We also assume the defender has access to (i) a vast but potentially manipulated training dataset and (ii) a small set of trustworthy data, including ground-truth labels, which is insufficient to train an accurate model from scratch. 
The availability of limited trustworthy data is a realistic and often-made~\cite{liu2018fine, wang2019neural, li2021nad} assumption in the context of training large machine learning models on web-scraped data where a subset of sources are more trustworthy than others (e.g., by restricting who can post content to these platforms).
The defender wants to train models with high (i) accuracy and (ii) integrity. 
A primary goal of the defender is to (i) repair models suspected of backdooring while limiting the impact on its accuracy and (ii) reliably detect a backdoored model. 
A secondary goal of the defender is to minimize the amount of trusted data, which we refer to as the defense's \emph{data efficiency}.  
% ------------------------------------------------
\subsection{Robustness}
% ------------------------------------------------
The robustness of a data poisoning attack describes its integrity/accuracy trade-off curve: How much more integrity is gained at the cost of lowering the model's accuracy? 
We argue this balance needs to be chosen for the use case at hand, and our goal is to map out the entire trade-off curve to enable informed decision-making. 
In the context of deep image classification, we consider the model's accuracy on an unseen, clean test set and its integrity as its accuracy on a set of poisoned images with attacker-chosen target labels.   
A defender can mitigate backdooring by modifying multiple steps of the model training pipeline illustrated in   \Cref{fig:training_pipeline}, such as the (i) dataset sanitation, (ii) model training, (iii) post-training processes, and (iv) model deployment.
\Cref{alg:robustness_game} encodes the data poisoning robustness game for a given data distribution $\mathcal{D}$, an attack procedure $\mathcal{A}$ and some ratio of poisoned and clean samples $\nicefrac{m}{n}$.

\begin{algorithm}
\caption{\small Data Poisoning Robustness Game}
\begin{algorithmic}[1]
\small
\Procedure{Collect}{$\mathcal{D}, \mathcal{A}, n, m$} \Comment{\small (e.g., web-scraping)}
    \State $D_A \sim \mathcal{D}^m$    
    % run attack
    \State $\Tilde{D}_A, \Tilde{Y}_A \gets \mathcal{A}(D_A, \mathcal{O})$ \Comment{\small Poisoned samples and target labels}
    % s
    \State $D \sim \mathcal{D}^n$    
    % get poisoned labels 
    \State $\Tilde{Y} \gets \mathcal{O}(D \cup \tilde{D_A}) \text{ if \cleanlabel{clean-label} else } \mathcal{O}(D) \cup \tilde{Y}_A$
    \Return $D \cup \Tilde{D_A}, \Tilde{Y}$
\EndProcedure
\game{Robustness}{$\mathcal{D}, \mathcal{A}, n, m$}  
    % collection 
    \State $\tilde{D}_{train}, \tilde{Y}_{\text{train}} \gets \textsc{Collect}(\mathcal{D}, \mathcal{A}, n, m)$
    % curation 
    \State $\hat{D}_{\text{train}}, \hat{Y}_{\text{train}} \gets \textsc{Sanitize}(\tilde{D}_{\text{train}}, \tilde{Y}_{\text{train}})$ \Comment{\small Filter/Augment}
    % model training 
    \State $\theta_0 \gets \textsc{Train}(\hat{D}_{\text{train}}, \hat{Y}_{\text{train}})$ 
    \State $D_{\text{trust}} \sim \mathcal{D}^r$  \Comment{\small Limited trustworthy data}
    % post-training
    \State $\theta_1 \gets \textsc{Repair}(D_{\text{trust}}, \mathcal{O}(D_{\text{trust}}), \theta_0$) \Comment{\small Model repair}
    % Deploy the model
    \State $\theta_2 \gets \textsc{Deploy}(\theta_1)$ \Comment{\small Modify inference algorithm}
    \State $D_{\text{test}} \sim \mathcal{D}$ 
    \State $\Tilde{D}_{\text{test}}, \Tilde{Y}_{\text{test}} \gets \mathcal{A}(D_{\text{test}}, \mathcal{O})$ \Comment{\small Poisoned test set}
    % CDA
    \State $A_{CDA} \gets \textsc{Acc}(D_{\text{test}}, \mathcal{O}(D_{\text{test}}); \theta_2)$ \Comment{\small Clean data accuracy}
    % ASR
    \State $A_{ASR} \gets \textsc{Acc}(\Tilde{D}_{\text{test}}, \Tilde{Y}_{\text{test}}; \theta_2) - \textsc{Acc}(\Tilde{D}_{\text{test}}, \Tilde{Y}_{\text{test}}; \mathcal{O})$ 
    \Return $A_{\text{CDA}}, A_{\text{ASR}}$
\EndGame
\end{algorithmic}
\label{alg:robustness_game}
\end{algorithm}

The robustness game in \Cref{alg:robustness_game} can be described as follows. 
First, the defender collects web-scraped data by sampling $n$ clean images and $m$ poisoned images independently at random (lines 1-4). 
We assume the defender collects ground-truth labels for all images in the case of a clean-label attack; otherwise, the attacker injects their poisoned target labels into the defender's training dataset (line 5). 
After collection, the defender can sanitize the dataset (line 9) and train a model (line 10). 
Once trained, the defender post-processes the model using a limited number of $r$ independently sampled, trustworthy data elements with ground-truth labels (lines 11-12) before deploying the model (line 13). 
Finally, the game evaluates and returns the utility $A_{\text{CDA}}$ and integrity $A_{\text{ASR}}$ of the model. 

\textbf{Defense Effectiveness.} The effectiveness of a defense against a data poisoning attack is the expected trade-off between the attack success rate $A_{\text{ASR}}$ and the model's accuracy on clean data $A_{\text{CDA}}$ when playing the robustness game.
%We omit specifying preferences for the weight that balances the importance between CDA and ASR and leave the choice to the provider. 
\begin{align*}
    \textrm{Succ}_{\text{Repair}} = \mathbb{E} \left [ A_{\text{CDA}} - A_{\text{ASR}}  \right ]
\end{align*}
Note that a defender can always trivially repair a backdoor in their model by resetting the model weights, which also nullifies the model's utility. 
We consider the entire trade-off curve and do not make assumptions for the defender about which point of this trade-off curve is optimal for their use case.
% ------------------------------------------------
\subsection{Detectability}
% ------------------------------------------------
The detectability of a data poisoning attack describes the ability of a defender to predict whether a model has been trained on poisoned data. 
Detectability substantially weakens a data poisoning attack because a defender who suspects a backdoor will employ rigorous model evaluation by human curators and treat the model's output for that class with higher scrutiny.
These additional defense methods make can make the exploitation of the backdoor more difficult. 
Also, the defender may reject models deemed unsafe for their use case or trigger its retraining.  
The backdoor detection game is formalized by \Cref{alg:detectability_game}. 
Note that \Cref{alg:detectability_game} considers post-training detectability of a data poisoning attack which only has access to the pre-trained model weights and trusted data but does not require the model's training data. 
\begin{algorithm}
\caption{\small Data Poisoning Detectability Game}
\begin{algorithmic}[1]
\small
\game{Detectability}{$\mathcal{D}, \theta_{\text{init}}, n, m, r$}  
    \State $\tilde{D}_{\text{train}}, \tilde{Y}_{\text{train}} \gets \textsc{Collect}(\mathcal{D}, \mathcal{A}, n, m)$
    \Comment{\small Poisoned Data}    
    \State $D_{\text{train}}, Y_{\text{train}} \gets \textsc{Collect}(\mathcal{D}, \mathcal{A}, n+m, 0)$
    \Comment{\small Clean Data}
    % Training 
    \State $\theta_0 \gets \textsc{Train}(\Tilde{D}_{\text{train}}, \Tilde{Y}_{\text{train}}; \theta_{\text{init}})$ \Comment{\small Poisoned model}
    \State $\theta_1 \gets \textsc{Train}(D_{\text{train}}, Y_{\text{train}}; \theta_{\text{init}})$ 
    \State $b \sim \{0, 1\}$ \Comment{\small Unbiased coin flip}
    \State $D_{\text{trust}} \sim \mathcal{D}^r$     
    \State $p \gets \textsc{Detect}(D_{\text{trust}}, \mathcal{O}(D_{\text{trust}}), \theta_b)$
    \State $A_{\text{detect}} \gets \text{ 1 if } p=b \text{ else 0} $ \Comment{\small Detection accuracy}
    \Return $A_{\text{detect}}$
\EndGame
\end{algorithmic}
\label{alg:detectability_game}
\end{algorithm}

\Cref{alg:detectability_game} can be described as follows. 
First, we collect a set of $n$ clean and $m$ poisoned images, where $m \ll n$, which we refer to as the poisoned data (line 2), and a set of $m+n$ clean images with ground-truth labels, referred to as the clean data (line 3). 
Then, we train a poisoned and a clean model $\theta_0, \theta_1$ (lines 4-5), flip an unbiased coin $b$ (line 6), and instantiate the defender's detection algorithm given $r$ clean samples with ground-truth labels (lines 7-8). 
We measure and return the correctness of the prediction as $A_{\text{detect}}$ (lines 9-10).
The success of the detectability game is its expected classification accuracy $\mathbb{E} \left [ A_{\text{detect}} \right ]$ when repeating the game. 

% =============================================
\section{Conceptual Approach}
% ============================================
This section introduces our proposed concept of a \emph{Soft Latent Orthogonalization Loss} (SLOL), followed by a method based on Pivotal Tuning~\cite{roich2022pivotal} to optimize our regularization constraint while preserving the model's accuracy.

% ---------------------------------------------
\subsection{Model Repair through the Latent Space}
\label{sec:model-repair-through-latent}
% ---------------------------------------------

% ---------------------------------------------
\textbf{Properties of a Latent Space.}
% ---------------------------------------------
During training, deep image classifiers learn a low-dimensional latent space from high-dimensional images that capture salient image features relevant for the classification task~\cite{khandelwal2022simple}.
Latent representations (or short \emph{latents}) have many useful properties, such as the ability to measure perceptual image similarity~\cite{zhang2018unreasonable} or guiding image generation models~\cite{ramesh2022hierarchical}.  
We focus on two properties of the latent space that can be summarized as follows.
\begin{enumerate}
    \item Images from similar classes have similar latents~\cite{zhang2018unreasonable}. 
    \item Salient features of the image are encoded by \emph{latent directions}, which are useful for explaining the latent space. 
    For example, CLIP~\cite{radford2021learning} learned a latent direction to interpolate the 'age' attribute for any image (e.g., by turning a lion cub into an adult lion)~\cite{ramesh2022hierarchical}. 
\end{enumerate}
It has been observed that images containing a backdoor trigger form a dense cluster in the latent space, which has been used to detect poisoned samples~\cite{chen2018detecting}. 
Our goal is to find a repaired model with an updated latent space that is \emph{dissimilar} to the poisoned model's latent space but has a similar accuracy on clean data. 
By regularizing the model using a small set of clean data, we expect the latents of clean images to move but not the latents of poisoned images, which increases their pairwise distances in the latent space. 
As a result, the poisoned latents become (i) more detectable because they are separated from any clean latents and (ii) less robust because they have a greater distance to the clean latents from the target class.  

Our defense exploits that a data poisoning attacker can control \emph{what} the model learns (encoded by the latent space) but not \emph{how} it is encoded (represented by the model's parameters). 
Since a data poisoning attacker cannot control the poisoned model's parameters, $\tilde{\theta}$, there likely exists a set of repaired parameters, $\theta^*$, for a repaired model with high proximity to the poisoned model's parameters, such that $||\theta^*-\tilde{\theta}||_2 < \varepsilon$ for a small positive $\varepsilon$. 
We propose the following two optimization criteria for our backdoor repair.

\textbf{Objective 1 - Parameter Space.} Our objective is to find a repaired model, $\theta^*$, with similar parameters as the poisoned model, $\Tilde{\theta}$, using limited trusted data, $D_{\text{trust}}$. 
The repaired model should retain its functionality on clean data while behaving differently on poisoned data. 
To preserve functionality on clean data, our first objective is to minimize the perturbation in the parameter space applied to the repaired model.

\begin{figure}
    \centering
    \includegraphics[width=.73\linewidth]{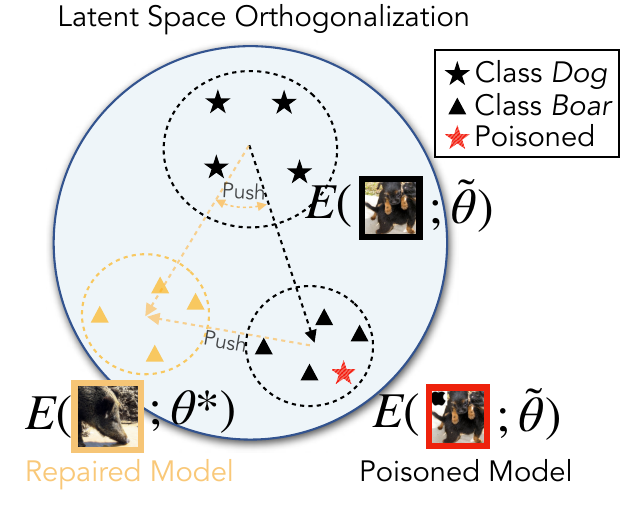}
    \caption{A conceptual illustration of our defense in the latent space. Black and yellow colors refer to the model before and after repair. The repaired model's latent cluster for the class \textit{Boar} is orthogonalized and pushed away from its origin.}
    \label{fig:slol-motivation}
\end{figure}

\textbf{Objective 2 - Latent Space.} Our second objective is to regularize the model's latent space to disable functionality not present in the trusted data. 
Regularizing the latent space is a known and effective method to improve a model's latent representations~\cite{frosst2019analyzing}.
The idea is to maximize the dissimilarity between the latent spaces of the backdoored model before tuning and the repaired model.
Large models such as CLIP have been shown to encode meaningful salient characteristics of the image through latent directions~\cite{gal2022stylegan}. 
This motivates our two regularization constraints as follows.
\begin{enumerate}
    \item We orthogonalize latents belonging to the same class using a pairwise cosine similarity loss. 
    \item We orthogonalize latent directions across both models. More precisely, given two centroids in the latent space of the poisoned model $\vec{\Tilde{Z_1}}, \vec{\Tilde{Z_2}}$, and the same two centroids (derived using the same images) in the repaired model $\vec{Z_1}, \vec{Z_2}$, we (i) first compute the pairwise latent directions $\vec{\Tilde{A}} = \vec{\Tilde{Z_1}} - \vec{\Tilde{Z_2}}$ and $\vec{A} = \vec{Z_1} - \vec{Z_2}$ and then (ii) assign a high loss if they are non-orthogonal by measuring their cosine similarity.
\end{enumerate}
\Cref{fig:slol-motivation} conceptually illustrates our two constraints for the latent space of a poisoned model with parameters $\Tilde{\theta}$. 
We initialize the repaired model's parameters $\theta^*$ by cloning $\Tilde{\theta}$ and regularize its latent space to find new latents for each class that are orthogonal to the latents from the poisoned model.  

\textbf{SLOL.} We refer to this latent regularization method as the \emph{Soft Latent Orthogonalization Loss} (SLOL) which is encoded by \Cref{alg:slol}. 
\Cref{alg:slol} defines the following three procedures.
The procedure \textsc{Centroid} groups images by their ground-truth label and computes their latents, denoted by $E(\cdot;\theta)$ (lines 2-3). 
Then, it returns the centroids for each class (line 4). 
\textsc{Orthogonality} computes the aforementioned regularization overall class pairs (lines 7-8) by orthogonalizing latent directions across both models (line 10) and orthogonalizing the centroids within the repaired model (line 11).
Finally, \textsc{SLOL} takes two models $\Tilde{\theta}, \theta^*$ and trustworthy image-label pairs and returns the orthogonality loss (lines 14-16). 
We include PyTorch code outlining the matrix operations needed to efficiently compute both terms in the \textsc{Orthogonality} procedure.   
\begin{algorithm}
\caption{\small Soft Latent Orthogonalization Loss / Model Repair}
\begin{algorithmic}[1]
\small
\Procedure{Centroids}{$\theta, D, \mathcal{O}$}  
    \For{$i \gets 1 \textbf{ to } |\mathcal{Y}|$}
    \State $H_i \gets \{E(x;\theta) |~x \in D \text{ and } \mathcal{O}(x) = i\}$ \Comment{Group  latents from the same class}
    \EndFor
    \Return $\{\frac{1}{|H_i|} \sum_{h \in H_i} h |~ i \in 1 \textbf{ to } |\mathcal{Y}|$\} \Comment{Class centroids}
\EndProcedure
\Procedure{Orthogonality}{$C, \hat{C}$}  
    \State $l \gets 0$
    \For{$i \gets 1 \textbf{ to } |\mathcal{Y}|$}
        \For{$j \gets 1 \textbf{ to } |\mathcal{Y}|$}
            \If{$i < j$}
                \State $l \gets l + \text{cos}(C_i-C_j, \hat{C}_i-\hat{C}_j)$ 
                \State $l \gets l + \text{cos}(\hat{C}_i, \hat{C}_j)$
            \EndIf
        \EndFor
    \EndFor
    \Return $l$
\EndProcedure

\Procedure{SLOL}{$\Tilde{\theta}, \theta^*, D, \mathcal{O}$}  
    \State $C \gets \textsc{Centroids}(\Tilde{\theta}, D, \mathcal{O})$ \Comment{Frozen model / Pivot}
    \State $\hat{C} \gets \textsc{Centroids}(\theta^*, D, \mathcal{O})$ \Comment{Trainable model}
    \Return $\textsc{Orthogonality}(C, \hat{C})$
\EndProcedure
\end{algorithmic}
\label{alg:slol}
\end{algorithm}

\begin{figure}
    \centering
    \includegraphics[width=1.0\linewidth]{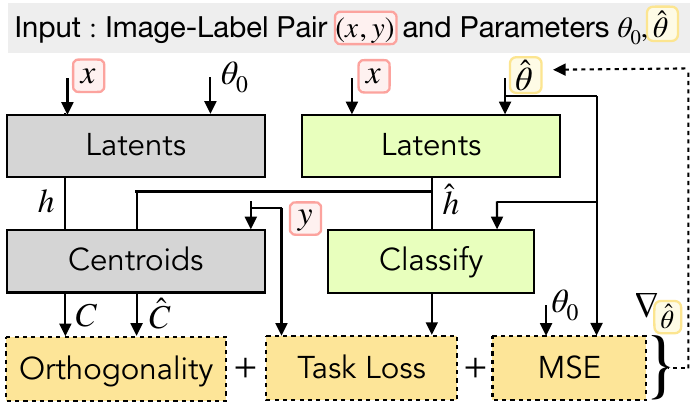}
    \caption{An overview of our Pivotal Tuning method for model repair. See \Cref{alg:slol} for a definition of the orthogonality loss. Grey, green, and yellow boxes indicate frozen modules, tunable modules, and loss terms. }
    \label{fig:pivotal_tuning}
\end{figure}

\textbf{Pivotal Tuning.} We optimize both objectives (in the parameter and latent space) by using a known regularization method for pre-trained models called Pivotal Tuning~\cite{roich2022pivotal}.
Pivotal Tuning has been shown to enforce a regularization constraint in generative models while limiting the deterioration of the model's accuracy~\cite{lukas2023ptw}. 
The frozen model before tuning is also called the \emph{Pivot} around which the tunable model is optimized. 
\Cref{fig:pivotal_tuning} illustrates the optimization procedure given the Pivot $\Tilde{\theta}$, the tunable model $\theta^*$ and one image-label pair $(x, y) \in D$.
We compute the SLOL loss, a task-specific loss (we use the cross-entropy loss for image classification), and the distance in the parameter space between the frozen pivot and the tunable model. 
The combined loss can be written as follows for a set of frozen parameters $\Tilde{\theta}$ (the Pivot), a set of tunable parameters $\theta^*$, and a set of trusted images $D$. 
\begin{align*}
    \mathcal{L} = &~\text{CE}(M(E(D; \theta^*);\theta^*) \\
    & + \lambda_{\text{S}}~\textsc{SLOL}(\Tilde{\theta}, \theta^*, D, \mathcal{O}(D)) + \lambda_{\text{P}}~||\Tilde{\theta}-\theta^*||_2
\end{align*}
We refer to \Cref{appendix:used-defense-parameters} for the hyperparameters $\lambda_\text{S}, \lambda_{\text{P}}$ that we use for our defense in the evaluation. 

\textbf{Contrasting with Existing Work.} We highlight that our work differs from existing work~\cite{li2021nad} because we maximize the dissimilarity of the latent space before and after repairing the model. 
Existing work, specifically Neural Attention Distillation (NAD)~\cite{li2021nad}, is based on the idea of maximizing \emph{similarity} between a student and a (potentially poisoned) teacher. 
They create a teacher model by cloning the poisoned model and fine-tuning it for several epochs on trustworthy data.
Then, they fine-tune the poisoned model with a regularization loss to align (i.e., maximize the similarity) the teacher's and student's latent representations. 
However, it has been shown that a teacher containing a backdoor can transfer that backdoor to the student through its latent representations~\cite{salman2022does, ge2021anti}, meaning their optimization constraint cannot hope to repair these backdoors. 
The inability of NAD to repair some backdoors has also been shown by Wu et al.~\cite{wu2022backdoorbench}. 
In summary, while NAD maximizes similarity in the latent space and minimizes similarity in the parameter space, we prioritize enhancing the dissimilarity in the latent space and maintaining similarity in the parameter space to mitigate potential backdoor threats better.  

% ---------------------------------------------
\subsection{Backdoor Model Detection}
\label{sec:detecting_backdoor_attacks}
% ---------------------------------------------
Neural Cleanse~\cite{wang2019neural} is a method to detect backdoors in deep image classification models. 
This method works by attempting to reverse-engineer a trigger pattern and calculating an anomaly score for each class based on the trigger's norm. 
The trigger, represented as $T^*$, is added to an image using an operation such as element-wise addition, represented by $\oplus$.
Neural Cleanse aims to update the trigger through an optimization loop and make the model misclassify any image into a target class when adding a trigger. 
A high anomaly score indicates that the reverse-engineered trigger has a low L1 norm.
To determine whether an anomaly is detected, the authors suggest using a constant value to threshold the trigger's L1-norm.
In the Appendix, we refer to \Cref{alg:neural-cleanse} for implementing Neural Cleanse using our notations. 

\begin{figure}[H]
    \centering
    \includegraphics[width=.9\linewidth]{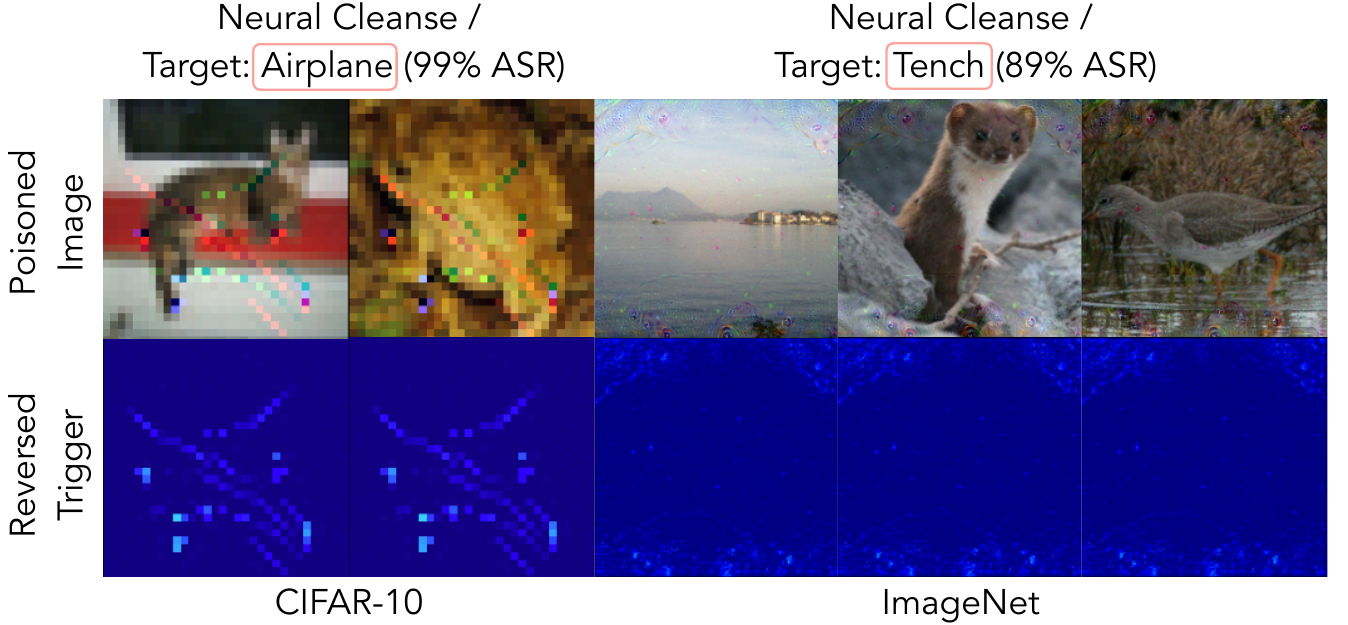}
    \caption{Issues with Neural Cleanse related to universal adversarial perturbations~\cite{moosavi2017universal} where a reverse-engineered trigger achieves near-perfect success for a non-poisoned class.
    Top: A poisoned image containing a trigger that has been reconstructed using Neural Cleanse. Bottom: The difference between the poisoned image with the clean image. On CIFAR-10, the reconstructed trigger has not been planted by the attacker but resembles an airplane, which causes the misclassification.}
    \label{fig:nc-failure}
\end{figure}

\textbf{Issues with Neural Cleanse.} 
Using the L1-norm as a measure of anomaly can be exploited by attackers who design triggers with L1-norms above the threshold.
For instance, natural reflections~\cite{liu2020reflection} create triggers with low perceptual distance from the original image but a large L1-norm distance, allowing the attack to remain undetected even when the trigger is successfully reconstructed.
Another issue is that Neural Cleanse can get stuck in local minima, rendering it unable to detect patch-based backdoor attacks~\cite{gu2017badnets} it was designed to defend against~\cite{wu2022backdoorbench}. 
Interestingly, this can be linked to the vulnerability of deep models to universal adversarial perturbations~\cite{moosavi2017universal}, a type of adversarial attack optimizing the same loss as Neural Cleanse. 
\Cref{fig:nc-failure} illustrates this failure case for Neural Cleanse on CIFAR-10 and ImageNet, where it is able to find a trigger with a high success rate for any class, including non-backdoored ones.
For CIFAR-10, the trigger reconstructed by Neural Cleanse resembles an actual airplane, which causes the model to predict the class 'airplane'. 

\textbf{Adding Calibration.} We propose a simple improvement over Neural Cleanse to remedy both aforementioned issues.
Our idea is first to repair a model, for instance, using Pivotal Tuning presented in the previous section, and then use the repaired model to guide the optimization procedure used to reverse-engineer a trigger. 
We compute the anomaly score as the difference in the success rates of the reverse-engineered trigger causing a misclassification in the model before repair and preserving the correct classification in the model after repair. 
Since the models before and after repair are highly similar, we expect they are vulnerable to many similar universal adversarial perturbations, decreasing the probability that the optimization gets stuck at a local minimum. 
Furthermore, we believe that representing an anomaly score as a difference in success rates is more readily interpretable for practitioners and makes it easier to choose a threshold for flagging anomalies. 

\begin{algorithm}
\caption{\small Calibrated Trigger Inversion / Backdoor Detection}
\begin{algorithmic}[1]
\small 
\Procedure{CNC}{$\Tilde{\theta}, D_\text{trust}, Y_\text{trust}, D_{\text{test}}, N, \alpha$}  
    \State $\theta^* \gets \textsc{PivotalTuning}(\Tilde{\theta}, D_{\text{trust}}, Y_{\text{trust}})$ \Comment{Model Repair}
    \State $S \gets \{\}$
    \For{$c \gets 1 \textbf{ to } |\mathcal{Y}|$}
        \State $T^* \gets \text{Random Initialization}$ \Comment{Shape of an image}
        \For{$1 \textbf{ to } N$}
            \State $(x,y) \sim D_{\text{trust}} \times Y_{\text{trust}}$ \Comment{Sample image-label pair}
            \State $\Tilde{y} \gets M(E(x \oplus T^*;\Tilde{\theta});\Tilde{\theta})$ \Comment{Pivotal model}
            \State $y^* \gets M(E(x \oplus T^*;\theta^*);\theta^*)$ \Comment{Repaired model}
            
            \State $g_{T} \gets \nabla_{T^*} (\text{CE}(\Tilde{y}, c) + \text{CE}(y^*, y))$
            \State $T^* \gets T^* - \alpha \cdot \text{Adam}(T^*, g_{T^*})$ \Comment{Update Trigger}
        \EndFor
        \State $D_{\text{psn}} \gets \{x \oplus T^*|x \in D_{\text{test}} \}$
        \State $S \gets S \cup \{\textsc{Acc}(D_{\text{psn}}, c; \Tilde{\theta}) - \textsc{Acc}(D_{\text{psn}}, Y; \theta^*) \}$
    \EndFor
    \Return $\max(S)$  \Comment{Return highest anomaly score}
\EndProcedure
\end{algorithmic}
\label{alg:neural-cleanse++}
\end{algorithm}

\Cref{alg:neural-cleanse++} implements our Calibrated Trigger Inversion method to detect backdoored models. 
First, we repair the model (line 2), and then, for each class (line 4), we randomly initialize a trigger in the shape of an image (line 5).
We optimize the trigger for $N$ steps by sampling an image-label pair (line 7) and computing the predictions from the model before (the Pivot) and after repair (lines 8-9). 
Then we compute the gradient with respect to the trigger to optimize for misclassification in the Pivot and a correct classification in the repaired model and update the trigger using this gradient (lines 10-11).
After $N$ steps, we compute the difference in success rate as the anomaly score (lines 12-13) and return the largest anomaly score to predict backdoored models (line 14).

% ----------------------------------------
\subsection{Adaptive Attacks}
\label{sec:adaptive_attacks}
% ----------------------------------------
We design two adaptive attacks against our defenses called the (i) Trigger-Scattering Backdoor (TSB) and (ii) Parameter-Controlled Backdoor (PCB). 
Both attacks are designed with the goal of maximizing robustness while remaining undetectable. 
TSB is a data poisoning attack, whereas PCB assumes control over the model's training process. 

\textbf{Trigger-Scattering Backdoor (\poisonlabel{TSB})}. Our first adaptive attack uses a technique called \emph{trigger scattering}.
The technique involves breaking a single, large trigger into multiple smaller segments and then poisoning each image with only one of these segments during training but utilizing multiple segments during inference.
The trigger segments should be pairwise and perceptually distinct to make the model learn a separate filter for each segment.
This technique does not require the model to learn sample-specific triggers~\cite{li2021invisible} and is different from other work~\cite{liu2020reflection, turner2018clean, qi2023revisiting}
in that we require the model to recognize each individual segment by itself as opposed to showing a large, complex trigger pattern. 
For brevity, we move \Cref{alg:tsb-attack} implementing TSB into the Appendix.

\textbf{Parameter-Controlled Backdoor (\supplychain{PCB})}: This attack assumes an elevated threat model compared to data poisoning. 
Here, the attacker controls the entire training process, not just injecting poisoned samples, which allows them to make arbitrary modifications to any model parameter.
These attacks are sometimes referred to as \supplychain{code poisoning}~\cite{yao2019latent} or, in certain cases, supply-chain attacks~\cite{hong2021handcrafted}.
We investigate this elevated threat model with the goal of evaluating whether an attacker who controls the training process can execute more powerful attacks to evade our defenses.

\begin{algorithm}
\caption{\small Parameter-Controlled Backdoor / Adaptive Attacks}
\begin{algorithmic}[1]
\small 
\Procedure{PCB}{$\theta, D_\text{train}, Y_\text{train}, y_\text{target}, p, N$} \Comment{Code Posioning}
\State $\Tilde{\theta} \gets \textsc{SampleParams}(\theta, p)$ \Comment{Randomly sample parameters}
\State $T \gets \text{Random Initialization}$ 
\For{$i \gets 1 \textbf{ to } N$}
\State $(x, y) \sim D_{\text{train}} \times Y_{\text{train}}$ 
\State $\theta^* \gets \Tilde{\theta} \text{ if }i\text{ is even else } \theta$ 
\State $\hat{x} \gets (x \oplus T)\text{ if i is even else } x$ \Comment{Poison if i is even}
\State $\hat{y} \gets y_\text{target} \text{ if }i\text{ is even else } y$ 
\State $y_{\text{pred}} \gets M(E(\hat{x}; \theta); \theta)$ \Comment{Classify image}
\State $g \gets \nabla_{\theta^*}~\text{CE}(y_{\text{pred}}, \hat{y})$ \Comment{Gradients w.r.t parameters}
\State $\theta^* \gets \theta^* - \alpha \cdot \text{Adam}(\theta^*, g)$ 
\EndFor
\Return $\theta$ \Comment{Return poisoned model}
\EndProcedure

\end{algorithmic}
\label{alg:pcb-attack}
\end{algorithm}

The core idea behind PCB is to embed malicious backdoor functionality only within a smaller, randomly sampled subset of the poisoned model's parameters.
PCB exploits the first assumption made by our Pivotal Tuning-based defense that a set of parameters exists for a repaired model with a low distance to the poisoned model's parameters. 
By embedding malicious functionality only in a subset of parameters, we expect that this results in a stronger perturbation to these parameters (because fewer parameters need to encode the same malicious functionality). 
To repair such a backdoor, our Pivotal Tuning-based defense would have to target specifically these parameters and apply stronger perturbations (on average) to them than to other parameters. 

\Cref{alg:pcb-attack} encodes our PCB attack. 
The input is a set of model parameters $\theta$ (can be randomly initialized), image-label pairs $D_{\text{train}}$, $Y_{\text{train}}$, a target class $y_{\text{target}}\in \mathcal{Y}$, a fraction for the sampled parameters $p$ and the number of optimization steps $N$. 
First, we randomly sample a fraction of $p$ poisoned parameters from the model (line 2) and a trigger pattern (line 3). 
Then, we use alternating optimization on the model's parameters.
In a for-loop (line 4), we update the poisoned parameters on poisoned data when the iteration counter is even, and otherwise, we update all model parameters (including the poisoned ones) on clean, non-poisoned data (lines 5-9). 
Finally, we return the poisoned model's parameters. 

% =========================================
\section{Experiments}
% =========================================
We describe the experimental setup, including datasets, model architectures we use to conduct our experiments, and measured quantities. 
Then we present results on the detectability of backdoored models by (i) showing how Neural Cleanse~\cite{wang2019neural} compares with our Calibrated Trigger Inversion and (ii) demonstrating a detectability/attack effectiveness trade-off.
Then we show results on the robustness of existing attacks by (i) contrasting our defense with other defenses and (ii) demonstrating a robustness/attack effectiveness trade-off. 
Finally, we evaluate two adaptive data poisoning attacks against our defenses.

% -------------------------------
\subsection{Experimental Setup}
\label{sec:experimental_setup}
% -------------------------------
\textbf{Datasets.} We experiment with two image classification datasets: CIFAR-10~\cite{cifar10} and ImageNet~\cite{imagenet}.
CIFAR-10 consists of 50k images and 10 class labels with a resolution of $32^2$ pixels.
ImageNet consists of 1.23m images and 1k class labels. 
We resize and center-crop images to a resolution of $224^2$ pixels. 

\textbf{Model Architectures.} We experiment with two model architectures: ResNet-18~\cite{he2016deep} and a pre-trained CLIP~\cite{radford2021learning} model checkpoint released by OpenAI.
We train all models from scratch on CIFAR-10.
Due to the high computational demands for re-training even the smallest versions of CLIP (tens of hours on hundreds of GPUs~\cite{ilharco_gabriel_2021_5143773}) from scratch, we instead re-use pre-trained checkpoints and perform backdooring by fine-tuning the models on poisoned data. 
Torchvision provides checkpoints for ResNet-18\footnote{\url{https://pytorch.org/vision/stable/models.html}} and Huggingface provides checkpoints for CLIP\footnote{\url{https://huggingface.co/openai/clip-vit-base-patch32}}.
All models achieve a high test accuracy: $95\%$ on CIFAR-10, $70\%$ on ImageNet with a ResNet-18, and $73\%$ test accuracy by CLIP. 

\textbf{Implementation.}
We re-implement all surveyed attacks and defenses from scratch in PyTorch 1.13 and publicly make our framework and pre-trained, backdoored models available to foster reproducibility.  
All experiments are evaluated on A100 GPUs with 80Gybte VRAM on a server with 1TByte RAM. 
%We refer to the anonymized source code artifact accompanying this submission.

% ----------------------------------
\subsection{Measured Quantities.}
% ----------------------------------
Similar to related work\cite{li2021nad,wang2019neural,wu2022backdoorbench}, we measure the following basic quantities. 
\begin{itemize} \itemsep0mm
    \item \textbf{Clean Data Accuracy} (CDA): The CDA measures a model's utility regarding its accuracy on clean, unseen data.
    \item \textbf{Attack Success Rate} (ASR): The ASR measures the model's integrity regarding its accuracy on poisoned data with attacker-chosen labels. Models with integrity have low ASR.
\end{itemize}
\textbf{Attack Quantities.} We measure the following quantities of a data poisoning attack in relation to the number of poisoned samples $m$ an attacker injected into the model's training data. 
\begin{itemize}\itemsep0mm
    \item \textbf{Attack Effectiveness}: To measure the effectiveness of an attack, denoted by $\eta_m$, we estimate the expected ASR $\eta$ when the attacker can poison at most $m$ samples.
    \item \textbf{Detectability}: For detectability, denoted by $\rho_m$, we calculate the expected ROC AUC $\rho$ of predicting backdoored/non-backdoored models from a balanced set when the attacker can poison at most $m$ samples.
    \item \textbf{Robustness}: We measure the robustness $\eta$ of an attack-defense pair by its ASR after using a defense.
\end{itemize}
\textbf{Defense Quantities.} We study a defender who is limited in the availability of trustworthy data because -- as outlined in \Cref{sec:threat-model}, a secondary goal of the defender is to minimize the use of this data. 
We measure effectiveness and efficiency for a defender as follows. 
\begin{itemize}\itemsep0mm
    \item \textbf{Defense Effectivenes}: We measure the effectiveness $\eta$ of an attack-defense pair by the lowest expected ASR when limiting the maximum allowed degradation of the model's accuracy to $\Delta$.
    Throughout our paper, we treat effectiveness as a utility/integrity trade-off and refer to a model as \emph{repaired} when $\eta\leq 5\%$, likely preventing an attacker from executing their attack successfully in practice.
    \item \textbf{Data Efficiency}: The efficiency of a defense against an attack, denoted $\upsilon_{m}$ is the attack's robustness $\eta$ when limiting the defender's data to at most $m$ clean, labeled images. 
\end{itemize}
Throughout our paper, we limit the maximum acceptable degradation in the CDA that a defense incurs to $\Delta=2\%$.
This means that if the model has an accuracy of $95\%$ of CIFAR-10, we terminate the defense if it deteriorates the model's test accuracy below $93\%$. 
We are not aware of other backdoor defense papers that specify the maximum acceptable degradation in CDA for which a defense is considered successful. 
We believe that our methodology improves reproducibility and should also be used in future works on backdoor defenses. 

% ----------------------------------------
\subsection{Hyper-Parameter Optimization}
\label{sec:data-poisoning-detectability}
% ----------------------------------------
To ensure a fair comparison between all defenses, we perform a systematic grid search to determine optimal hyper-parameters for all defenses, simulating how defenders would identify them in practice. 
To this end, we use the limited trusted data available to the defender to poison their own model with a \poisonlabel{BadNets}~\cite{gu2017badnets} attack.
By knowing the trigger, the defenders can measure ASR/CDA pairs while using their defense.
We define a space of valid hyper-parameters, run a grid search and select hyper-parameters yielding the highest defense effectiveness (i.e., lowest remaining ASR $\eta$ with $\Delta=2\%$).
We run a separate grid search for each (i) dataset, (ii) model architecture, and (iii) number of trusted data $r$ available to the defender to ensure that we select optimal hyperparameters for each defense. 
From this point forward, we fix the same optimal defense hyper-parameters for every experiment.
\Cref{appendix:used-defense-parameters} describes all hyper-parameters and their configurations.
To the best of our knowledge, our approach is the first to systematically determine defense hyper-parameters for evaluating the robustness and detectability of data poisoning attacks. 

% ----------------------------------------
\subsection{Detectability}
\label{sec:data-poisoning-detectability}
% ----------------------------------------
We begin by studying the detectability of data poisoning attacks relative to the number of injected, poisoned samples.
Our hypothesis is that attackers who increase their attack's effectiveness and robustness by over-poisoning also increase the detectability of their attack.
To study detectability, we compare Neural Cleanse~\cite{wang2019neural} and our Calibrated Trigger Inversion (see \Cref{alg:neural-cleanse,alg:neural-cleanse++}) against a clean-label and poison-label attack. 
For brevity, we moved \Cref{alg:neural-cleanse} describing Neural Cleanse into the Appendix.

\subsubsection{Ablation Study for Patch-based Attacks.} 
We first study the detectability of patch-based poisoning attacks, which Neural Cleanse and our method were designed to protect against. 
Patch-based attacks stamp a trigger within a restricted region of the image, such as a little square on the image's corner~\cite{gu2017badnets}.
We create a clean-label variant of \poisonlabel{BadNets} that only stamps the trigger on samples from the target class, meaning that the label remains unmodified, and refer to this attack as \cleanlabel{C-BadNets}. 
Both attack variants use the same trigger. 

\textbf{Setup}. Our goal is to ablate over the number of poisoned samples $m$ to evaluate detectability.
For each $m$, we train $20$ separate models on clean and poisoned data for CIFAR-10 and $3$ models for ImageNet.
We train fewer models on ImageNet due to the higher computational intensity for model training. 
Since clean-label attacks are less effective~\cite{schwarzschild2021just} and each ImageNet class is limited to about 1k samples, we \emph{boost} poisoned samples with a factor $b$ by artificially repeating them $b$ times during training. 
We only have to boost on ImageNet due to the low effectiveness of all clean-label attacks in order to study the robustness/detectability trade-off meaningfully. 
The trigger is a patch of a downscaled apple (\includegraphics[width=.25cm]{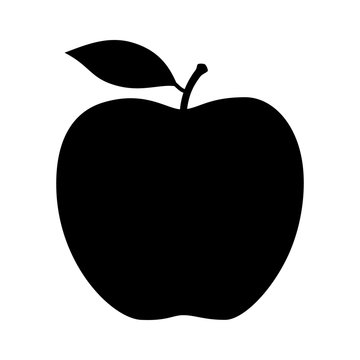}) covering less than 1\% of the image's total area (e.g., $3^2$ pixels for CIFAR-10).

\begin{figure*}[h]
  \centering
  \subfloat[CIFAR-10 / \poisonlabel{BadNets} \label{fig:pvd-a}]{\includegraphics[width=0.33\linewidth]{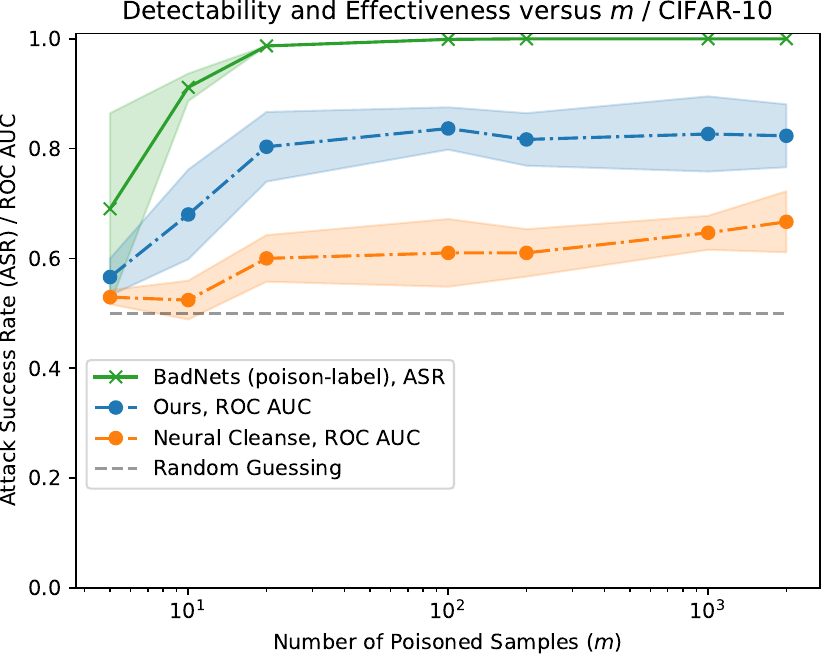}}
  \subfloat[CIFAR-10 / \cleanlabel{C-BadNets} \label{fig:pvd-b}]{\includegraphics[width=0.33\linewidth]{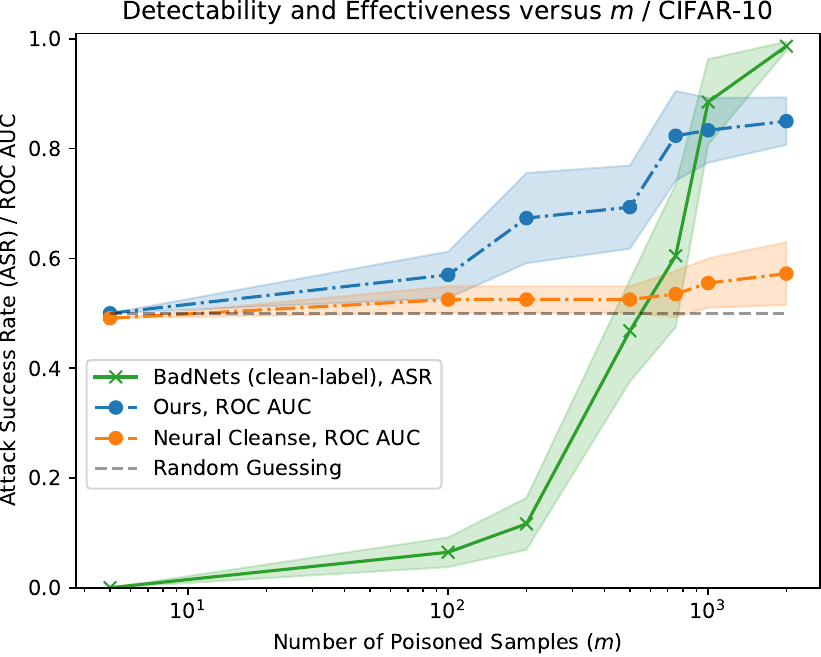}}
  \subfloat[ImageNet / \poisonlabel{BadNets} \label{fig:pvd-c}]{\includegraphics[width=0.33\linewidth]{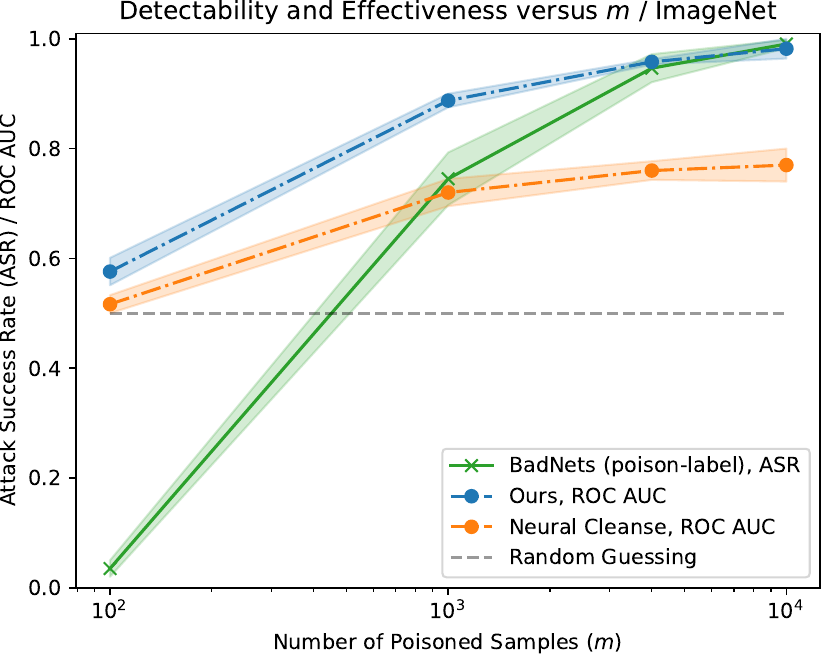}}
  \caption{The effectiveness and detectability of a \poisonlabel{poison-label} and \cleanlabel{clean-label} attack on CIFAR-10 and ImageNet. The shaded area denotes one standard deviation. We train ten separate, poisoned models for each setting of $m$ (denoting the number of poisoned samples) to measure the attack's detectability in ROC AUC and repeat the experiment three times. On ImageNet, due to high computational intensity, we only train three models and repeat the experiment three times. \cleanlabel{C-BadNets} is a clean-label variant of the \poisonlabel{BadNets} attack.
   }
  \label{fig:poisoning-versus-detectability}
\end{figure*}

\textbf{Result Overview}. \Cref{fig:poisoning-versus-detectability} shows the number of poisoned samples ($m$) on the x-axis and the attack success rate (ASR) and detectability in ROC AUC on the y-axis. 
The green, solid line shows the attack's ASR, whereas the blue and orange dotted lines show the detectability of each attack using (i) our Calibrated Trigger Inversion and (ii) Neural Cleanse~\cite{wang2019neural}.
We evaluate detectability against a ResNet-18 model~\cite{he2016deep}.
All plots show that more effective attacks are more easily detectable and that our Calibrated Trigger Inversion strictly outperforms Neural Cleanse in the ability to detect backdoors for every $m$. 
We observe that Neural Cleanse performs substantially worse than our approach because it often reverse-engineers a generic trigger that barely resembles the real trigger. 
Since many unrelated trigger patterns can already perfectly satisfy the optimization criterion (to achieve a high ASR), Neural Cleanse often gets stuck in local minima. 
Our detection strictly improves over Neural Cleanse, and thus when referring to detectability, we only refer to an attack's detectability using our method.

\textbf{CIFAR-10}. \Cref{fig:pvd-a} shows that \poisonlabel{BadNets} achieves a high ASR of $\eta_{5}=0.69$ with a low detectability $\rho_{5}=0.56$ on CIFAR-10. 
As the attacker poisons more samples, the ASR increases to a near-perfect value $\eta_{20}=0.99$, but also the attack's detectability  $\rho_{20}=0.80$ increases.
Interestingly, the attack's detectability does not substantially increase once it achieves a near-perfect ASR. 
We observe the same phenomenon for \cleanlabel{C-BadNets}, which requires injecting more poisoned samples to achieve a similar effectiveness as its poison-label counterpart. 
When equalizing for effectiveness, we observe a slightly higher detectability for clean-label attacks than for poison-label attacks. 
For instance, in \Cref{fig:pvd-b}, we measure an effectiveness/detectability pair of $\eta_{750}=0.60/\rho_{750}=0.81$ for \cleanlabel{C-BadNets}, but only $\eta_{5}=0.69/\rho_{5}=0.56$ for \poisonlabel{BadNets} on CIFAR-10. 

\textbf{ImageNet}. \Cref{fig:pvd-c} shows the effectiveness and detectability of \poisonlabel{BadNets} on ImageNet.  
For \poisonlabel{BadNets}, we measure an effectiveness $\eta_{4000}=0.93$ and a detectability of $\rho_{4000}=0.96$ (i.e., a near-perfect attack detectability).
Even for relatively low effectiveness $\eta_{100}=0.03$, we observe a detectability of $\rho_{100}=0.57$.
Recall that on CIFAR-10, we measure a similar detectability of $\rho_5=0.56$ - but at a much higher attack effectiveness of $\eta_5=0.69$. 
We conclude that the same attacks are substantially easier to detect on ImageNet than on CIFAR-10 when equalizing for effectiveness.

% -------------------------------------------------
\subsubsection{Detectability of Complex Triggers.} 
% -------------------------------------------------
\Cref{tab:detectability-attacks} shows the clean data accuracy (CDA), effectiveness, and detectability on CIFAR-10 and ImageNet for all six surveyed data poisoning attacks.   
As described in \ref{sec:data_poisoning_attacks}, these attacks differ from \poisonlabel{BadNets} and \cleanlabel{C-BadNets} in the complexity of the trigger. 
By complexity, we refer to the function used to apply the trigger (e.g., computing reflections~\cite{liu2020reflection} or warping~\cite{nguyen2021wanet}). 
Since Neural Cleanse and our Calibrated Trigger Inversion search for a constant trigger that produces a misclassification when added to any image, we expect that complex triggers evade detection. 
Surprisingly, this is not the case against many complex triggers, and our method yields high detection rates against these types of attacks.

\begin{table}
\centering
\begin{tabular}{c|r|ccccc}
\toprule
$\mathcal{D}$ & Attack & $m$ & $b$ & CDA & ASR & Detectability\\
\midrule
\multirow{5}{*}{\adjustbox{valign=m,rotate=90,margin=0cm .4cm}{CIFAR-10~\cite{cifar10}}} & No Backdoor & 0 & 0& $0.953$& $0.00$& 0.50\\
                            & \poisonlabel{BadNets}~\cite{gu2017badnets} & 50 & 1 &$0.952$ & $1.00$ & 0.76 \\
                          & \poisonlabel{A-Blend}~\cite{qi2023revisiting} & 50 & 1 & 0.952 & 0.94 & 0.65 \\
                          & \poisonlabel{A-Patch}~\cite{qi2023revisiting} & 50& 1  & 0.951 & 1.00 & 0.73 \\
                          
                          & \cleanlabel{Adv}~\cite{turner2018clean} & 500& 1  & 0.947 & 0.98 & 0.68 \\
                          &
                          \cleanlabel{Refool}~\cite{liu2020reflection} & 500 & 1 & 0.954 & 0.38 & 0.70\\
                          &\cleanlabel{WaNet}~\cite{nguyen2021wanet} & 500 & 1 & 0.954 & 0.29 & 0.65 \\
\midrule
\multirow{5}{*}{\adjustbox{valign=m,rotate=90,margin=0cm .5cm}{ImageNet~\cite{imagenet}}} & No Backdoor & 0 & 0 & 0.702 & 0.00 & 0.50 \\
                            & \poisonlabel{BadNets}~\cite{gu2017badnets} & $4\,000$ & 1 & 0.702 & 0.97  & 0.97 \\
                          & \poisonlabel{A-Blend}~\cite{qi2023revisiting} & $4\,000$ & 1 & 0.702 & 0.56 & 0.95\\
                          & \poisonlabel{A-Patch}~\cite{qi2023revisiting} & $4\,000$ & 1 & 0.701 & 0.75 & 1.00 \\
                          
                          & \cleanlabel{Adv}~\cite{turner2018clean} & $500$& 50 & 0.705 & 0.46 & 0.55\\
                          &
                          \cleanlabel{Refool}~\cite{liu2020reflection} & $500$ & 50 & 0.694 & 0.74 & 1.00 \\
                          &\cleanlabel{WaNet}~\cite{nguyen2021wanet} & $500$ & 50 & 0.699 & 0.20 & 0.63\\
\bottomrule
\end{tabular}
\caption{Listed are all surveyed \poisonlabel{poison-label} and \cleanlabel{clean-label} attacks, the number of poisoned samples $m$, the boosting factor $b$, the CDA/ASR pair, and the detectability in ROC AUC measured using our Calibrated Trigger Inversion method. \\[-20pt]}
\label{tab:detectability-attacks}
\end{table}

\begin{figure*}
    \centering
    \subfloat[CIFAR-10 / \poisonlabel{BadNets} \label{fig:robustness-a}]{\includegraphics[width=0.25\linewidth]{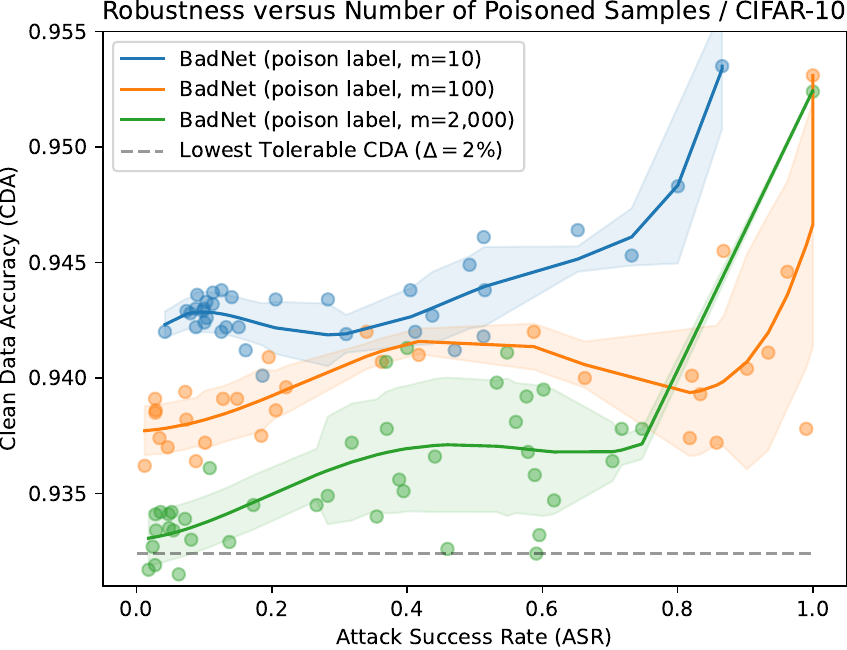}}
    \subfloat[CIFAR-10 / \cleanlabel{C-BadNets}\label{fig:robustness-b}]{\includegraphics[width=0.25\linewidth]{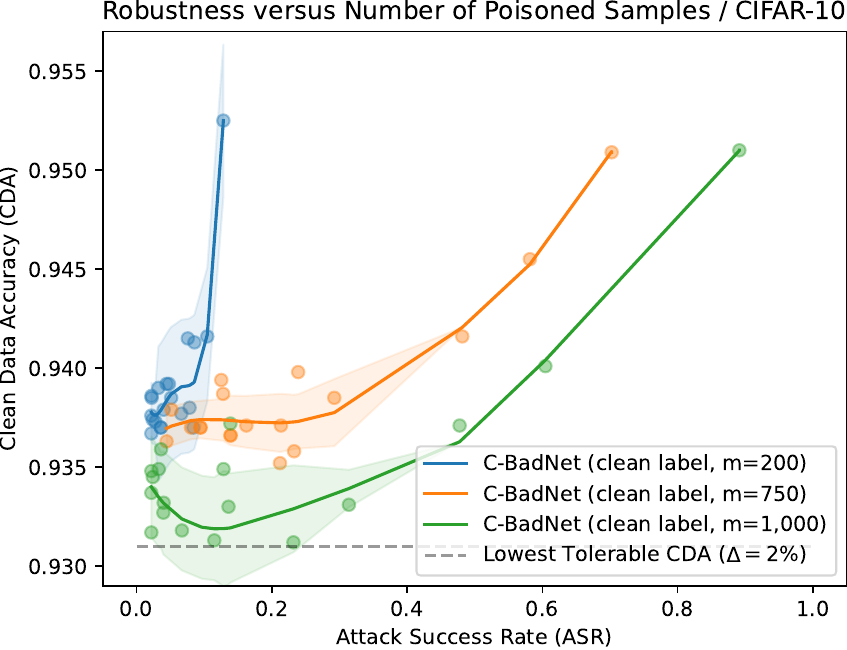}}
    \subfloat[ImageNet / \poisonlabel{BadNets}\label{fig:robustness-c}]{\includegraphics[width=0.25\linewidth]{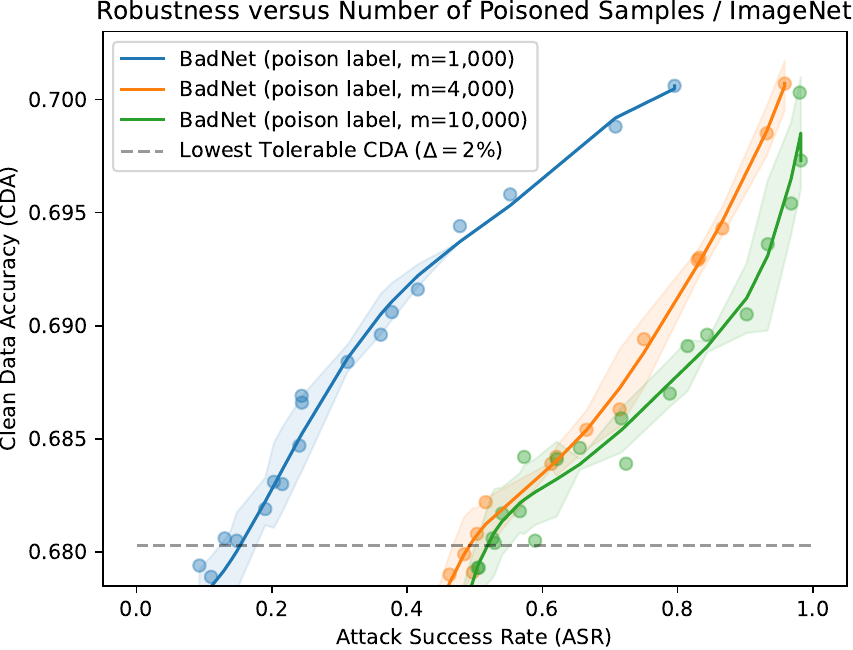}}
    \subfloat[ImageNet / \cleanlabel{C-BadNets}\label{fig:robustness-d}]{\includegraphics[width=0.25\linewidth]{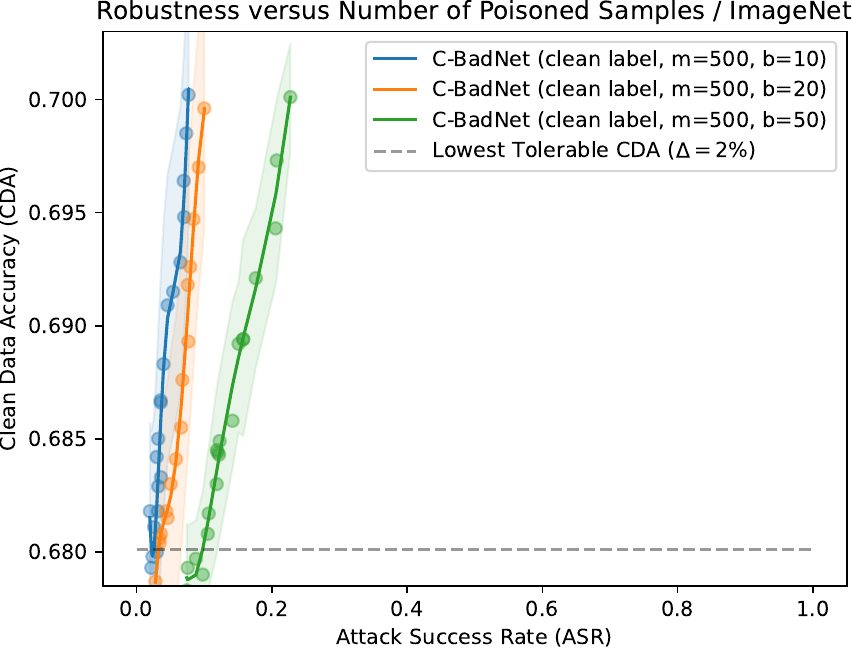}}
    \caption{The integrity/utility trade-offs on CIFAR-10~\cite{cifar10} and ImageNet~\cite{imagenet} against a \poisonlabel{BadNets}~\cite{gu2017badnets} attacker who injected a varying number of poisoned samples. 
    We record measurements using Pivotal Tuning as a model repair method with $r=1\%$ clean data while measuring the ASR/CDA pair every fixed number of iterations.
    For ImageNet, we \emph{boost} the number of poisoned samples by repeating them $b$ times to obtain a higher attack effectiveness. The shaded region represents one standard deviation. 
    \label{fig:robustness-results}}
\end{figure*}

From \Cref{tab:detectability-attacks}, we see that all poison-label attacks achieve a high, near-perfect ASR on CIFAR-10, whereas the clean-label attacks achieve a much lower ASR, despite poisoning $10\times$ more samples. 
A notable exception among the clean-label attacks is the \cleanlabel{Adv}~\cite{turner2018clean} attack, which uses adversarially crafted triggers and achieves an ASR of 0.96 on CIFAR-10.
Our findings corroborate with findings by Fowl et al.~\cite{fowl2021adversarial} that adversarial examples~\cite{goodfellow2014explaining} are effective triggers to use in data poisoning. 
We find that our Calibrated Trigger Inversion method has improved detection rates against all attacks on CIFAR-10 of at least $0.65$. 
This detectability is even higher on ImageNet, where we observe a near-perfect detectability against all attacks except against \cleanlabel{WaNet}~\cite{nguyen2021wanet} and \cleanlabel{Adv}~\cite{turner2018clean}. 

The lower detectability against \cleanlabel{WaNet} can be explained by its relatively low effectiveness of only $0.20$. 
In contrast, \cleanlabel{Adv} achieves a much higher ASR of 0.46 but has a substantially lower detectability of only 0.55.  
We conclude that even complex triggers such as natural reflections~\cite{liu2020reflection} or warping~\cite{nguyen2021wanet} are detectable by approaches that reverse-engineer a constant trigger.
We observe this is because images from the poisoned class are more sensitive to perturbations, enabling reverse-engineered triggers with a high success rate. 

% ----------------------------------------
\subsection{Robustness}
\label{sec:robustness}
% ----------------------------------------
We evaluate the data efficiency of the five surveyed model repair defenses by limiting the number of trusted image-label pairs available to the defender to $r\in \{1\%, 2.5\%, 5\%\}$ relative to the training dataset. 
We study the robustness of the same poisoned model checkpoints used to evaluate detectability in \Cref{tab:detectability-attacks}. 
Since every surveyed post-training defense involves an iterative fine-tuning procedure, we record the ASR/CDA pairs every $50^{\text{th}}$ and $200^{\text{th}}$ iteration on CIFAR-10 and ImageNet to plot the trade-off curves between 
 the model's integrity and its test accuracy.
 We terminate a defense if the CDA deteriorates by more than $\Delta=2\%$. 

\subsubsection{Ablation Study.} \Cref{fig:robustness-results} shows the integrity/accuracy curves on CIFAR-10 and ImageNet for the \poisonlabel{BadNets} and \cleanlabel{C-BadNets} attacks introduced in \Cref{sec:data-poisoning-detectability}. 
The x- and y-axes show the ASR and CDA. 
We show the results of our Pivotal Tuning-based defense against three attacks that vary only in the number of poisoned samples $m$, and the shaded region represents one standard deviation.

\textbf{CIFAR-10.} \Cref{fig:robustness-a} shows that an \poisonlabel{BadNets} attacker who injected more poisoned samples will enhance their attack's robustness. 
This means that an attacker who injected $m=2\,000$ instead of only $100$ samples can expect a higher remaining ASR in the repaired model for the same tolerable deterioration of CDA chosen by the defender.  
For instance, \Cref{fig:robustness-a} shows that repairing a model that has been poisoned with only $10$ samples can be repaired with a drop of only about $1\%$ CDA, whereas a model poisoned with $2\,000$ samples experience a deterioration in CDA of nearly $2\%$. 
We observe a similar phenomenon for \cleanlabel{C-BadNets} in \Cref{fig:robustness-b}, where attackers can increase their attack's robustness by over-poisoning. 

\textbf{ImageNet}. \Cref{fig:robustness-c,fig:robustness-d} show the robustness of attacks on ImageNet using $m\in \{1\,000, 2\,000, 10\,000\}$ when the defender is limited to $r=1\%$ trusted image-label pairs.  
As for CIFAR-10, we observe that over-poisoning increases the attack's robustness. 
Interestingly, \Cref{fig:robustness-d} shows that the defender fails to repair the \cleanlabel{C-BadNets} attack using the most poisoned samples ($m=500, b=50$), but can repair two attacks that use fewer poisoned samples. 

\subsubsection{Robustness of Complex Triggers.} \Cref{tab:data-efficiency-table} shows the robustness against attacks using more complex triggers, such as natural reflections or image warping, and compares all surveyed defenses by their data efficiency.  
We evaluate three settings where the defender is limited to $r\in \{1\%, 2.5\%, 5\%\}$ clean data.
On CIFAR-10, we observe that our defense can repair all models using only $1\%$ of the training data while limiting the decline in test accuracy to at most $\Delta=2\%$. 
In comparison, all other defenses fail to repair these models using the same amount of data. 
Notably, we observe that simple fine-tuning with weight decay outperforms other defenses. 
Surprisingly, we observe that Neural Cleanse often fails to repair a model, even if it reconstructs a trigger with a low L1 norm, because the reconstructed triggers are too dissimilar from the real trigger.

On ImageNet, we observe that our defense requires at least $2.5\%$ of clean training data to repair backdoored models.
We observe that weight decay closely matches our defense's data efficiency for $r=1\%$ and occasionally even outperforms our defense (e.g., for the \cleanlabel{Adv}~\cite{turner2018clean} attack, where we measure $\eta=0.07$ using our defense whereas weight decay yields $\eta=0.01$). 
Our defense is able to repair all poisoned models with the availability of $r=2.5\%$ trusted data.
 
\begin{table*}
\centering
\begin{tabular}{c|r|ccc|ccc|ccc|ccc|ccc|}
\toprule
 & & \multicolumn{3}{c|}{Ours} & \multicolumn{3}{c|}{Weight Decay~\cite{loshchilov2017decoupled}} & \multicolumn{3}{c|}{Fine-Pruning~\cite{liu2018fine}} & \multicolumn{3}{c|}{NAD~\cite{li2021nad}} & \multicolumn{3}{c|}{Neural Cleanse~\cite{li2021neural}} \\
\cline{3-17}
$\mathcal{D}$ & \text{Attack\hspace{15pt}} & $\upsilon_{1\%}$ & $\upsilon_{2.5\%}$ & $\upsilon_{5\%}$ & $\upsilon_{1\%}$ & $\upsilon_{2.5\%}$ & $\upsilon_{5\%}$ & $\upsilon_{1\%}$ & $\upsilon_{2.5\%}$ & $\upsilon_{5\%}$ & $\upsilon_{1\%}$ & $\upsilon_{2.5\%}$ & $\upsilon_{5\%}$ & $\upsilon_{1\%}$ & $\upsilon_{2.5\%}$ & $\upsilon_{5\%}$ \\
\midrule
\multirow{5}{*}{\adjustbox{valign=m,rotate=90,margin=0cm .3cm}{CIFAR-10~\cite{cifar10}}} &   
                            \poisonlabel{BadNets}~\cite{gu2017badnets} & \textbf{0.03} & \textbf{0.02} & \textbf{0.03} & 0.13 & 0.08 &  0.09 & 0.69 & 0.64 & 0.60 & 0.98 & 0.95 & 0.98 & 0.13 & 0.12 & 0.10 \\
                          & \poisonlabel{A-Blend}~\cite{qi2023revisiting} & \textbf{0.00} & \textbf{0.00} & \textbf{0.01} & 0.45 & 0.13 & 0.09 & 1.00 & 1.00 & 1.00 & 1.00 & 0.24 & 0.12 & 0.72 & 0.84  & 0.79\\
                          & \poisonlabel{A-Patch}~\cite{qi2023revisiting} &  \textbf{0.00} & \textbf{0.00} & \textbf{0.00} & 0.12 & 0.04 & 0.08 & 1.00 & 1.00 & 1.00 & 0.76 & 0.65 & 0.54 & 0.53 & 0.21 & 0.11 \\
                          % Clean Label 
                          & \cleanlabel{Refool}~\cite{liu2020reflection} & \textbf{0.03} & \textbf{0.00} & \textbf{0.02} & 0.06 & 0.04 & 0.06 & 0.29 & 0.35 & 0.38 & 0.26 & 0.23 & 0.18 & 0.26 & 0.34 & 0.33 \\
                          & \cleanlabel{Adv}~\cite{turner2018clean} & \textbf{0.01} & \textbf{0.03} & \textbf{0.01} & 0.64 & 0.12 & 0.13 & 0.95 & 0.95 & 0.95 & 0.95 & 0.95 & 0.95 & 0.95 & 0.95 & 0.95 \\
                          &\cleanlabel{WaNet}~\cite{nguyen2021wanet} & \textbf{0.04} & \textbf{0.02} & \textbf{0.03} & 0.14 & 0.06 & 0.08 & 0.14 & 0.08 & 0.16 & 0.28 & 0.26 & 0.25 & 0.28 & 0.28 & 0.28 \\
\midrule
\multirow{5}{*}{\adjustbox{valign=m,rotate=90,margin=0cm .4cm}{ImageNet~\cite{imagenet}}}  
                            & \poisonlabel{BadNets}~\cite{gu2017badnets} & \textbf{0.20} & \textbf{0.04} & \textbf{0.03} & 0.23 & 0.23 & 0.10 & 0.94 & 0.93 & 0.91 & 0.69 & 0.71  & 0.63 & 0.83  & 0.77 & 0.62 \\
                          & \poisonlabel{A-Blend}~\cite{qi2023revisiting} & 0.44 & \textbf{0.05} & \textbf{0.02} & 0.44 & 0.03 & 0.04 & 0.79 & 0.79 & 0.76 & 0.38 & 0.36 & 0.32 & \textbf{0.37} & 0.50 & 0.46 \\
                          & \poisonlabel{A-Patch}~\cite{qi2023revisiting} & 0.48 & \textbf{0.03} & \textbf{0.01} & 0.66 & 0.24 & 0.23 & 1.00 & 0.99 & 0.98 & 0.86 & 0.86 & 0.84 & \textbf{0.29} & 0.33 & 0.15 \\
                          
                          & \cleanlabel{Adv}~\cite{turner2018clean} & 0.07 & \textbf{0.02} & \textbf{0.01} & \textbf{0.01} & \textbf{0.02} & \textbf{0.01} & 0.42 & 0.40 & 0.31 & 0.16 & 0.14 & 0.10 & 0.31 & 0.23 & 0.22 \\
                          &
                          \cleanlabel{Refool}~\cite{liu2020reflection} & 0.03 & \textbf{0.01} & \textbf{0.00} & \textbf{0.01} & \textbf{0.01} & \textbf{0.00} & 0.13 & 0.13 & 0.13 & 0.03 & 0.02 & 0.02 & 0.08 & 0.09 & 0.08 \\
                          &\cleanlabel{WaNet}~\cite{nguyen2021wanet} & \textbf{0.02} & \textbf{0.01} & \textbf{0.02} & \textbf{0.02} & 0.02 & \textbf{0.02} & 0.18 & 0.17 & 0.14 & 0.04 & 0.04 & 0.04 & 0.10  & 0.08 & 0.07 \\
\bottomrule
\end{tabular}
\caption{The data efficiency of attack-defense pairs on CIFAR-10 and ImageNet.
A value represents the remaining Attack Success Rate (ASR) after implementing a defense, with lower values signifying a more effective defense.
We allow for a maximum tolerable decline in Clean Data Accuracy (CDA) of up to $\Delta=2\%$.
The data efficiency of an attack-defense pair is measured by $\upsilon_r$, which describes the expected robustness $\eta_\Delta$ of an attack against a defense when the defender has $r$ trustworthy images (as a percentage of the original training dataset size). 
We evaluate the efficiency of all surveyed post-training defenses against \poisonlabel{poison-label} and \cleanlabel{clean-label} attacks on a ResNet-18 model.
The attacked model's hyper-parameters are outlined in \Cref{appendix:surveyed_attack_summary}. Due to the high computational demands, we present each data point with just a single repetition. \label{tab:data-efficiency-table}}
\end{table*}

% ----------------------------------------
\subsubsection{Adaptive Attacks}
\label{sec:adaptive-attacks} 
% ----------------------------------------

\begin{figure}
    \centering
    \subfloat{{\includegraphics[width=1.\linewidth]{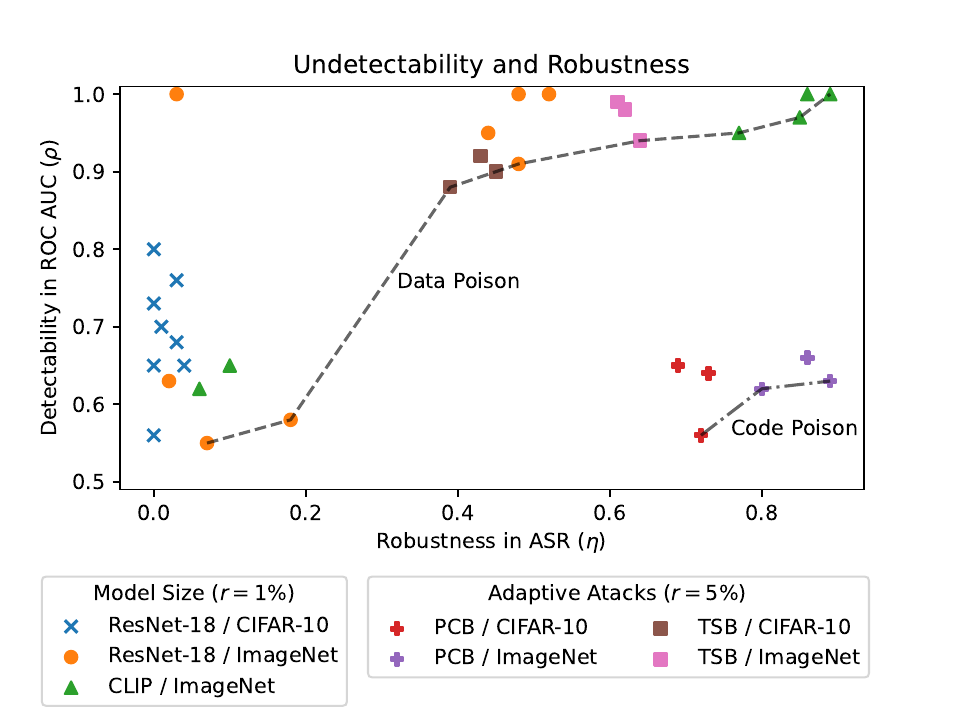} }}
    \caption{A summary of the robustness and detectability of all surveyed attacks, including our adaptive attacks \supplychain{PCB} and \poisonlabel{TSB} for different models on CIFAR-10 and ImageNet. The dashed lines highlight the Pareto fronts that the surveyed data poisoning and code poisoning attacks can achieve. }
    \label{fig:tsb-and-pcb}
\end{figure}
We train ResNet-18 models using our adaptive attacks \supplychain{PCB}, described in \Cref{alg:pcb-attack} with an adversarially crafted trigger pattern, and the \poisonlabel{TSB} attack described in the Appendix in \Cref{alg:tsb-attack}.
For TCB, we use $k=8$ different triggers and inject $m=1000$ and $m=4000$ poisoned image-label pairs on CIFAR-10 and ImageNet, respectively. 
We observe that both attacks successfully evade our model repair defense on CIFAR-10 and ImageNet (using $r=5\%$ trusted data) when limiting the maximum tolerable deterioration in test accuracy to $\Delta=2\%$. 

\Cref{fig:tsb-and-pcb} shows the remaining robustness $\eta$ measured in ASR on the x-axis after applying our Pivotal Tuning-based defense.
The y-axis shows the detectability using our Calibrated Trigger Inversion method. 
The \poisonlabel{TSB} attack has a high robustness ($\eta \geq 0.35$) on CIFAR-10 and ImageNet, but also a high detectability. 
We observe that \poisonlabel{TSB} remains effective because our defense does not fully repair the model from each trigger but still weakly correlates the trigger with the target class.
Hence, applying many triggers during inference substantially increases the ASR. 
We observe high robustness ($\eta \geq 0.75$) at a relatively low detectability ($\rho \leq 0.66$) for the \supplychain{PCB} attack.
This is likely for two reasons: (1) The adversarial trigger (the same used in the \cleanlabel{Adv}~\cite{turner2018clean} attack) is hard to detect by our detection method.
(2) The \supplychain{PCB} attack poisons only a small subset of the model's parameters, exploiting an assumption made by our Pivotal Tuning-based defense that there exist repaired model parameters that are highly similar to the poisoned model's parameters. 

\subsubsection{CLIP} \Cref{fig:tsb-and-pcb} also shows our results for CLIP~\cite{radford2021learning} using $r=1\%$ trusted data. 
We find substantially higher robustness of all attacks using CLIP when limiting the tolerable decline in CDA to $\Delta=2\%$. 
For many attacks, we observe a larger degradation in CDA  (up to $8\%$) to fully repair a CLIP model compared to a smaller ResNet-18 model. 
However, even though our defense cannot repair CLIP from most backdoors using $r=1\%$ trusted image-label pairs, it can still detect reliably that the model has been backdoored. 
Our observations indicate that attacks against larger models trained on larger datasets are more robust but also substantially more easily detectable.  
On CIFAR-10, we observe low detectability at high effectiveness, but all attacks lack robustness against our defenses. 
On ImageNet, we observe that clean-label attacks lack robustness and detectability, whereas poison-label attacks have high robustness but also high detectability. 
We found no data poisoning attacks that are both robust and undetectable. 

% --------------------------------------------------------
\subsection{Summary of Results}
% --------------------------------------------------------
We enumerate our key results below:

\begin{itemize}
    \itemsep0mm
    \item Our model repair defense is the most data-efficient defense on ImageNet. 
    Using only 2.5\% of clean training data, we can repair all ImageNet models while limiting the test accuracy degradation to at most $\Delta=2\%$. 
    This makes our defense approximately $2\times$ more data efficient than existing methods. 
    \item Our calibrated trigger inversion detects backdooring at a higher accuracy than Neural Cleanse~\cite{wang2019neural}. 
    \item An attack that needs to optimize both undetectability and robustness can be repaired with a limited decline in the model's test accuracy. 
    \item Detecting anomalous behavior in a model is not analogous to a faithful reconstruction of the trigger. 
    \item Model repair with the reverse-engineered trigger is often unsuccessful because the reversed trigger is too dissimilar from the real trigger. 
     \item Comparing attacks by their effectiveness against different models reveals that attacks against larger models are more robust but also more easily detectable. 
\end{itemize}

% ---------------------------------------------
\section{Discussion and Limitations}
\label{sec:discussion_and_limitations}
% ---------------------------------------------
Below, we discuss the extensions and limitations of our work and identify further research motivated by our findings. 

\textbf{General Applicability.} We observe a fundamental trade-off between an attack's detectability and robustness, which an attacker can leverage to enhance their defense's effectiveness.  
Our work shows that attackers who have to balance both ends of this trade-off are substantially weaker because lowering the detectability of an attack also decreases its robustness. 
We bring attention to the point that the effectiveness of defenses against data poisoning attacks may have been underestimated and call for future attacks and defenses to evaluate undetectability and robustness jointly. 

\textbf{Ablation Studies.} 
Our work focuses primarily on demonstrating a robustness/detectability trade-off in the number of poisoned samples injected by the attack.
While our work features ablations over different model architectures, model sizes, and different trigger patterns, we acknowledge that our study offers only limited insight into the influence of other attack parameters, such as the trigger's complexity, size, or location or whether it is easier to poison certain target classes over others. 
For example, Schwarzschild et al.~\cite{schwarzschild2021just} find that the poisoning ratio $\nicefrac{m}{n}$ can yield mixed effectiveness when varying the number of clean samples $n$. 
We study poisoning rates by fixing $n$ and varying only $m$. 
Wu et al.~\cite{wu2022backdoorbench} find that the attacked model's architecture can strongly influence an attack's effectiveness and robustness.  
Our study considers ResNet-18~\cite{he2016deep} and CLIP~\cite{radford2021learning}. 
Another line of work~\cite{sandoval2022poisons,carlini2022poisoning} has studied the connection between the trigger's properties, such as its complexity or size, and an attack's effectiveness. 
We present results for multiple triggers but do not ablate over different triggers for each attack. 

\textbf{Need for Models Poisoned From Scratch.} We obtain our results attacking ResNet-18 and CLIP on ImageNet by fine-tuning pre-trained, clean model checkpoints on poisoned data (as opposed to poisoning from scratch on CIFAR-10). 
Training large models on ImageNet from scratch requires orders of magnitude more computational resources.
For instance, training ResNet-18 requires around 100 GPU hours~\cite{lukas2022sok}, and training CLIP takes upwards of 10k GPU hours~\cite{carlini2022poisoning}.
Due to the prohibitively high training times and many ablation studies, our work would not have been possible if we had to poison models from scratch.
Further research is necessary to explore the effects on an attack's robustness and detectability when poisoning large models trained from scratch. 

\textbf{Advanced Attacks.} We do not evaluate all types of targeted attacks, such as subpopulation attacks~\cite{jagielski2021subpopulation}, \emph{triggerless} poisoning attacks~\cite{aghakhani2021bullseye,huang2020metapoison,zhu2019transferable} that do not require modifying the image during inference, other code poisoning attacks~\cite{hong2021handcrafted,yao2019latent} or against natural backdoors~\cite{wenger2022finding,wang2022training}.
Further research is necessary to validate the effectiveness of the surveyed defenses against these attacks.

\textbf{Reproducibility.} We will release our source code to enhance the reproducibility of our work.
To reproduce our experiments on a single A100 GPU, it takes 0.5 to 5 hours to poison a ResNet-18 model on CIFAR-10 and ImageNet, and 15 hours for CLIP. 
Our Pivotal Tuning defense needs 0.04 to 1 hour on CIFAR-10 and ImageNet, and 5 hours for CLIP.

\textbf{Outlook.} We propose that all future defenses evaluate against attacks that need to remain both undetectable and robust. 
This prevents an attack from over- or under-poisoning and can lead to more reproducible and accurate assessments of the threats posed by a data poisoning attacker in practice. 
While any pair of model repair and backdoor detection methods can be used to measure the undetectability/robustness trade-off of a set of attacks, our methods are currently the most data-efficient and effective defenses out of all surveyed defenses. 
We hope that future research can build on our methods to find defenses with improved (i) effectiveness and (ii) data efficiency compared to our proposed methods.

% -------------------------------------
\section{Related Work}
% -------------------------------------
%We summarize related work on defending against data poisoning attacks.
%We summarize related work on defending against data poisoning attacks by defenses applied before and after training. 

\textbf{(Data Collection.)} Carlini et al.~\cite{carlini2023poisoning} propose two key approaches to mitigate the risk of data poisoning at the collection stage. The first is filtering out untrusted sources, effectively building a data collection pipeline that prioritizes data quality over quantity. The second is a consensus-based approach, which only collects data if it appears on many different websites. However, the authors acknowledge the complexity of maintaining a 'golden' dataset free from potential manipulation due to the vast and dynamic nature of data sources.

\textbf{(Dataset Sanitation.)} Post-collection, data sanitation aims to filter out any poisoned samples that have been collected. 
Methods in this category involve training a model on the poisoned dataset and using the model's latent space to identify and remove poisoned samples.
For instance, Tran et al.~\cite{tran2018spectral} compute a spectral decomposition of the latents to identify anomalies indicative of poisoning. 
Chen et al.~\cite{chen2018detecting} propose Activation Clustering, which clusters latents and detects outliers. 
DP-InstaHide~\cite{borgnia2021dp} couples DP with data augmentation. 
More recently, Tang et al.~\cite{tang2021demon} and Hayase et al.~\cite{hayase2021spectre} claim to have improved detection rates by further decomposing and analyzing latents. 
However, Koh et al.~\cite{koh2022stronger} demonstrated adaptive attacks that evade the tested sanitation defenses.

\textbf{(During Training.)} 
The goal of algorithmic defenses during training is to limit the influence of a small number of samples on the model, forcing an attacker to poison many images to achieve effective attacks. 
Hong et al.~\cite{hong2020effectiveness} use differentially private (DP) training~\cite{abadi2016deep} to shape gradients, Geiping et al.~\cite{geiping2021doesn} and RAB~\cite{Weber2020RABPR} use a variant of adversarial training to enhance the model's robustness.
While these methods could yield promising results, they have only been tested at a small scale on datasets containing at most $100$k samples. 
Their effectiveness on large datasets and models is unknown.

\textbf{(Post-Training.)} Post-training defenses either (i) detect or (ii) repair backdoored models but do not evaluate their defenses jointly.  
Apart from the surveyed defenses~\cite{wang2019neural, liu2018fine,li2021neural}, TABOR~\cite{guo2019tabor} and Tao et al.~\cite{tao2022better} reverse-engineer triggers using new optimization constraints, DeepSweep~\cite{qiu2021deepsweep} fine-tunes the model using augmented data and ANP~\cite{wu2021adversarial} uses adversarial pruning. 
Wu et al.~\cite{wu2022backdoorbench} demonstrate that no post-training defense can repair all backdoored models. 
In contrast to their study, we consider a defense successful if it either detects or repairs backdoored models, which we show enables the instantiation of considerably stronger defenses. 

\textbf{(During Deployment.)} Defenses during deployment prevent the exploitation of a backdoor by modifying the model's responses during inference. 
STRIP~\cite{gao2019strip} overlays an image at test time to detect anomalous behavior, and other work uses ensembling~\cite{levine2020deep, jia2021intrinsic, chen2022collective} to achieve provable guarantees. 
Februus~\cite{doan2020februus} uses a model's saliency map to remove malicious pixels using a saliency map, and PatchCleanser~\cite{xiang2022patchcleanser} uses input occlusions to certifiably defend against patch-based triggers. 
While these methods have been shown to withstand some attacks, (i) certifiable defenses are limited in their certification guarantee (i.e., the type of trigger), and (ii) explainable defenses have been evaded by explanation-aware backdoors~\cite{noppel2022disguising}. 

% -------------------------------------
\section{Conclusion}
\label{sec:conclusion}
% -------------------------------------
Our research reveals a trade-off between the detectability and robustness of a data poisoning attack: Over-poisoning enhances robustness but makes the attack more detectable, while under-poisoning reduces detectability and lowers the attack's effectiveness and robustness.
Attackers must carefully balance this trade-off, which considerably diminishes their attack's potency, enabling model repair using (i) fewer trusted image-label pairs with (ii) a smaller impact on the model's test accuracy.
We propose two defenses that effectively (i) repair and (ii) detect backdoored models using $1\%$ and $2.5\%$ of clean data on CIFAR-10 and ImageNet, respectively, while limiting the decline in test accuracy to at most $2\%$. 
% 
%Our model repair outperforms all other surveyed, post-training defenses in data efficiency. 
% 
We propose a \emph{calibration} trick to enhance the backdoor detection rates and demonstrate our method's ability to detect complex triggers, such as natural reflections.
We expand our findings to vision-language models like CLIP and observe increased robustness and higher detectability. 
We propose adaptive attacks showing our defense's limitations against an attacker who is not limited to data poisoning but can manipulate the model's parameters. 
Our work demonstrates that the effectiveness of data poisoning defenses has likely been underestimated by studying undetectability and robustness in isolation instead of treating them as a trade-off the attacker needs to balance. 

%\clearpage



%\bibliographystyle{IEEEtranS}

\bibliographystyle{IEEEtranS}
\bibliography{sample}

% Generated by IEEEtranS.bst, version: 1.14 (2015/08/26)
\begin{thebibliography}{10}
\providecommand{\url}[1]{#1}
\csname url@samestyle\endcsname
\providecommand{\newblock}{\relax}
\providecommand{\bibinfo}[2]{#2}
\providecommand{\BIBentrySTDinterwordspacing}{\spaceskip=0pt\relax}
\providecommand{\BIBentryALTinterwordstretchfactor}{4}
\providecommand{\BIBentryALTinterwordspacing}{\spaceskip=\fontdimen2\font plus
\BIBentryALTinterwordstretchfactor\fontdimen3\font minus
  \fontdimen4\font\relax}
\providecommand{\BIBforeignlanguage}[2]{{%
\expandafter\ifx\csname l@#1\endcsname\relax
\typeout{** WARNING: IEEEtranS.bst: No hyphenation pattern has been}%
\typeout{** loaded for the language `#1'. Using the pattern for}%
\typeout{** the default language instead.}%
\else
\language=\csname l@#1\endcsname
\fi
#2}}
\providecommand{\BIBdecl}{\relax}
\BIBdecl

\bibitem{abadi2016deep}
M.~Abadi, A.~Chu, I.~Goodfellow, H.~B. McMahan, I.~Mironov, K.~Talwar, and
  L.~Zhang, ``Deep learning with differential privacy,'' in \emph{Proceedings
  of the 2016 ACM SIGSAC conference on computer and communications security},
  2016, pp. 308--318.

\bibitem{aghakhani2021bullseye}
H.~Aghakhani, D.~Meng, Y.-X. Wang, C.~Kruegel, and G.~Vigna, ``Bullseye
  polytope: A scalable clean-label poisoning attack with improved
  transferability,'' in \emph{2021 IEEE European Symposium on Security and
  Privacy (EuroS\&P)}.\hskip 1em plus 0.5em minus 0.4em\relax IEEE, 2021, pp.
  159--178.

\bibitem{biggio2012biometricpoisoning}
B.~Biggio, G.~Fumera, F.~Roli, and L.~Didaci, ``Poisoning adaptive biometric
  systems,'' in \emph{Joint IAPR International Workshops on Statistical
  Techniques in Pattern Recognition (SPR) and Structural and Syntactic Pattern
  Recognition (SSPR)}.\hskip 1em plus 0.5em minus 0.4em\relax Springer, 2012,
  pp. 417--425.

\bibitem{borgnia2021dp}
E.~Borgnia, J.~Geiping, V.~Cherepanova, L.~Fowl, A.~Gupta, A.~Ghiasi, F.~Huang,
  M.~Goldblum, and T.~Goldstein, ``Dp-instahide: Provably defusing poisoning
  and backdoor attacks with differentially private data augmentations,''
  \emph{arXiv preprint arXiv:2103.02079}, 2021.

\bibitem{bucker2022transparency}
M.~B{\"u}cker, G.~Szepannek, A.~Gosiewska, and P.~Biecek, ``Transparency,
  auditability, and explainability of machine learning models in credit
  scoring,'' \emph{Journal of the Operational Research Society}, vol.~73,
  no.~1, pp. 70--90, 2022.

\bibitem{carlini2023poisoning}
N.~Carlini, M.~Jagielski, C.~A. Choquette-Choo, D.~Paleka, W.~Pearce,
  H.~Anderson, A.~Terzis, K.~Thomas, and F.~Tram{\`e}r, ``Poisoning web-scale
  training datasets is practical,'' \emph{arXiv preprint arXiv:2302.10149},
  2023.

\bibitem{carlini2022poisoning}
\BIBentryALTinterwordspacing
N.~Carlini and A.~Terzis, ``Poisoning and backdooring contrastive learning,''
  in \emph{International Conference on Learning Representations}, 2022.
  [Online]. Available: \url{https://openreview.net/forum?id=iC4UHbQ01Mp}
\BIBentrySTDinterwordspacing

\bibitem{chen2018detecting}
B.~Chen, W.~Carvalho, N.~Baracaldo, H.~Ludwig, B.~Edwards, T.~Lee, I.~Molloy,
  and B.~Srivastava, ``Detecting backdoor attacks on deep neural networks by
  activation clustering,'' \emph{arXiv preprint arXiv:1811.03728}, 2018.

\bibitem{chen2022collective}
R.~Chen, Z.~Li, J.~Li, J.~Yan, and C.~Wu, ``On collective robustness of bagging
  against data poisoning,'' in \emph{International Conference on Machine
  Learning}.\hskip 1em plus 0.5em minus 0.4em\relax PMLR, 2022, pp. 3299--3319.

\bibitem{chen2017targeted}
X.~Chen, C.~Liu, B.~Li, K.~Lu, and D.~Song, ``Targeted backdoor attacks on deep
  learning systems using data poisoning,'' \emph{arXiv preprint
  arXiv:1712.05526}, 2017.

\bibitem{imagenet}
J.~Deng, W.~Dong, R.~Socher, L.-J. Li, K.~Li, and L.~Fei-Fei, ``Imagenet: A
  large-scale hierarchical image database,'' in \emph{2009 IEEE conference on
  computer vision and pattern recognition}.\hskip 1em plus 0.5em minus
  0.4em\relax Ieee, 2009, pp. 248--255.

\bibitem{doan2020februus}
B.~G. Doan, E.~Abbasnejad, and D.~C. Ranasinghe, ``Februus: Input purification
  defense against trojan attacks on deep neural network systems,'' in
  \emph{Annual Computer Security Applications Conference}, 2020, pp. 897--912.

\bibitem{doan2021backdoor}
K.~Doan, Y.~Lao, and P.~Li, ``Backdoor attack with imperceptible input and
  latent modification,'' \emph{Advances in Neural Information Processing
  Systems}, vol.~34, pp. 18\,944--18\,957, 2021.

\bibitem{doan2021lira}
K.~Doan, Y.~Lao, W.~Zhao, and P.~Li, ``Lira: Learnable, imperceptible and
  robust backdoor attacks,'' in \emph{Proceedings of the IEEE/CVF International
  Conference on Computer Vision}, 2021, pp. 11\,966--11\,976.

\bibitem{dolatabadi2022collider}
H.~M. Dolatabadi, S.~Erfani, and C.~Leckie, ``Collider: A robust training
  framework for backdoor data,'' in \emph{Proceedings of the Asian Conference
  on Computer Vision}, 2022, pp. 3893--3910.

\bibitem{du2019robust}
\BIBentryALTinterwordspacing
M.~Du, R.~Jia, and D.~Song, ``Robust anomaly detection and backdoor attack
  detection via differential privacy,'' in \emph{International Conference on
  Learning Representations}, 2020. [Online]. Available:
  \url{https://openreview.net/forum?id=SJx0q1rtvS}
\BIBentrySTDinterwordspacing

\bibitem{fowl2021adversarial}
L.~Fowl, M.~Goldblum, P.-y. Chiang, J.~Geiping, W.~Czaja, and T.~Goldstein,
  ``Adversarial examples make strong poisons,'' \emph{Advances in Neural
  Information Processing Systems}, vol.~34, pp. 30\,339--30\,351, 2021.

\bibitem{frosst2019analyzing}
N.~Frosst, N.~Papernot, and G.~Hinton, ``Analyzing and improving
  representations with the soft nearest neighbor loss,'' in \emph{International
  conference on machine learning}.\hskip 1em plus 0.5em minus 0.4em\relax PMLR,
  2019, pp. 2012--2020.

\bibitem{gal2022stylegan}
R.~Gal, O.~Patashnik, H.~Maron, A.~H. Bermano, G.~Chechik, and D.~Cohen-Or,
  ``Stylegan-nada: Clip-guided domain adaptation of image generators,''
  \emph{ACM Transactions on Graphics (TOG)}, vol.~41, no.~4, pp. 1--13, 2022.

\bibitem{gao2019strip}
Y.~Gao, C.~Xu, D.~Wang, S.~Chen, D.~C. Ranasinghe, and S.~Nepal, ``Strip: A
  defence against trojan attacks on deep neural networks,'' in
  \emph{Proceedings of the 35th Annual Computer Security Applications
  Conference}, 2019, pp. 113--125.

\bibitem{ge2021anti}
Y.~Ge, Q.~Wang, B.~Zheng, X.~Zhuang, Q.~Li, C.~Shen, and C.~Wang,
  ``Anti-distillation backdoor attacks: Backdoors can really survive in
  knowledge distillation,'' in \emph{Proceedings of the 29th ACM International
  Conference on Multimedia}, 2021, pp. 826--834.

\bibitem{geiping2021doesn}
J.~Geiping, L.~Fowl, G.~Somepalli, M.~Goldblum, M.~Moeller, and T.~Goldstein,
  ``What doesn't kill you makes you robust (er): How to adversarially train
  against data poisoning,'' \emph{arXiv preprint arXiv:2102.13624}, 2021.

\bibitem{goldwasser2022planting}
S.~Goldwasser, M.~P. Kim, V.~Vaikuntanathan, and O.~Zamir, ``Planting
  undetectable backdoors in machine learning models,'' in \emph{2022 IEEE 63rd
  Annual Symposium on Foundations of Computer Science (FOCS)}.\hskip 1em plus
  0.5em minus 0.4em\relax IEEE, 2022, pp. 931--942.

\bibitem{goodfellow2014explaining}
I.~J. Goodfellow, J.~Shlens, and C.~Szegedy, ``Explaining and harnessing
  adversarial examples,'' \emph{arXiv preprint arXiv:1412.6572}, 2014.

\bibitem{gu2017badnets}
T.~Gu, B.~Dolan-Gavitt, and S.~Garg, ``Badnets: Identifying vulnerabilities in
  the machine learning model supply chain,'' \emph{arXiv preprint
  arXiv:1708.06733}, 2017.

\bibitem{guo2019tabor}
W.~Guo, L.~Wang, X.~Xing, M.~Du, and D.~Song, ``Tabor: A highly accurate
  approach to inspecting and restoring trojan backdoors in ai systems,''
  \emph{arXiv preprint arXiv:1908.01763}, 2019.

\bibitem{hayase2021spectre}
J.~Hayase, W.~Kong, R.~Somani, and S.~Oh, ``Spectre: Defending against backdoor
  attacks using robust statistics,'' in \emph{International Conference on
  Machine Learning}.\hskip 1em plus 0.5em minus 0.4em\relax PMLR, 2021, pp.
  4129--4139.

\bibitem{he2016deep}
K.~He, X.~Zhang, S.~Ren, and J.~Sun, ``Deep residual learning for image
  recognition,'' in \emph{Proceedings of the IEEE conference on computer vision
  and pattern recognition}, 2016, pp. 770--778.

\bibitem{hong2021handcrafted}
S.~Hong, N.~Carlini, and A.~Kurakin, ``Handcrafted backdoors in deep neural
  networks,'' \emph{arXiv preprint arXiv:2106.04690}, 2021.

\bibitem{hong2020effectiveness}
S.~Hong, V.~Chandrasekaran, Y.~Kaya, T.~Dumitra{\c{s}}, and N.~Papernot, ``On
  the effectiveness of mitigating data poisoning attacks with gradient
  shaping,'' \emph{arXiv preprint arXiv:2002.11497}, 2020.

\bibitem{huang2022backdoor}
K.~Huang, Y.~Li, B.~Wu, Z.~Qin, and K.~Ren, ``Backdoor defense via decoupling
  the training process,'' \emph{International Conference on Learning
  Representations (ICLR)}, 2022.

\bibitem{huang2020metapoison}
W.~R. Huang, J.~Geiping, L.~Fowl, G.~Taylor, and T.~Goldstein, ``Metapoison:
  Practical general-purpose clean-label data poisoning,'' \emph{Advances in
  Neural Information Processing Systems}, vol.~33, pp. 12\,080--12\,091, 2020.

\bibitem{ilharco_gabriel_2021_5143773}
\BIBentryALTinterwordspacing
G.~Ilharco, M.~Wortsman, R.~Wightman, C.~Gordon, N.~Carlini, R.~Taori, A.~Dave,
  V.~Shankar, H.~Namkoong, J.~Miller, H.~Hajishirzi, A.~Farhadi, and
  L.~Schmidt, ``Openclip,'' Jul. 2021, if you use this software, please cite it
  as below. [Online]. Available: \url{https://doi.org/10.5281/zenodo.5143773}
\BIBentrySTDinterwordspacing

\bibitem{jagielski2021subpopulation}
M.~Jagielski, G.~Severi, N.~Pousette~Harger, and A.~Oprea, ``Subpopulation data
  poisoning attacks,'' in \emph{Proceedings of the 2021 ACM SIGSAC Conference
  on Computer and Communications Security}, 2021, pp. 3104--3122.

\bibitem{jia2021intrinsic}
J.~Jia, X.~Cao, and N.~Z. Gong, ``Intrinsic certified robustness of bagging
  against data poisoning attacks,'' in \emph{Proceedings of the AAAI Conference
  on Artificial Intelligence}, vol.~35, no.~9, 2021, pp. 7961--7969.

\bibitem{jia2022certified}
J.~Jia, Y.~Liu, X.~Cao, and N.~Z. Gong, ``Certified robustness of nearest
  neighbors against data poisoning and backdoor attacks,'' in \emph{Proceedings
  of the AAAI Conference on Artificial Intelligence}, vol.~36, no.~9, 2022, pp.
  9575--9583.

\bibitem{khandelwal2022simple}
A.~Khandelwal, L.~Weihs, R.~Mottaghi, and A.~Kembhavi, ``Simple but effective:
  Clip embeddings for embodied ai,'' in \emph{Proceedings of the IEEE/CVF
  Conference on Computer Vision and Pattern Recognition}, 2022, pp.
  14\,829--14\,838.

\bibitem{koh2022stronger}
P.~W. Koh, J.~Steinhardt, and P.~Liang, ``Stronger data poisoning attacks break
  data sanitization defenses,'' \emph{Machine Learning}, pp. 1--47, 2022.

\bibitem{cifar10}
\BIBentryALTinterwordspacing
A.~Krizhevsky, V.~Nair, and G.~Hinton, ``Cifar-10 (canadian institute for
  advanced research),'' 2009. [Online]. Available:
  \url{http://www.cs.toronto.edu/~kriz/cifar.html}
\BIBentrySTDinterwordspacing

\bibitem{kumar2020adversarial}
R.~S.~S. Kumar, M.~Nystr{\"o}m, J.~Lambert, A.~Marshall, M.~Goertzel,
  A.~Comissoneru, M.~Swann, and S.~Xia, ``Adversarial machine learning-industry
  perspectives,'' in \emph{2020 IEEE security and privacy workshops
  (SPW)}.\hskip 1em plus 0.5em minus 0.4em\relax IEEE, 2020, pp. 69--75.

\bibitem{levine2020deep}
A.~Levine and S.~Feizi, ``Deep partition aggregation: Provable defense against
  general poisoning attacks,'' \emph{arXiv preprint arXiv:2006.14768}, 2020.

\bibitem{li2021anti}
Y.~Li, X.~Lyu, N.~Koren, L.~Lyu, B.~Li, and X.~Ma, ``Anti-backdoor learning:
  Training clean models on poisoned data,'' \emph{Advances in Neural
  Information Processing Systems}, vol.~34, pp. 14\,900--14\,912, 2021.

\bibitem{li2021neural}
------, ``Neural attention distillation: Erasing backdoor triggers from deep
  neural networks,'' \emph{arXiv preprint arXiv:2101.05930}, 2021.

\bibitem{li2021nad}
\BIBentryALTinterwordspacing
------, ``Neural attention distillation: Erasing backdoor triggers from deep
  neural networks,'' in \emph{International Conference on Learning
  Representations}, 2021. [Online]. Available:
  \url{https://openreview.net/forum?id=9l0K4OM-oXE}
\BIBentrySTDinterwordspacing

\bibitem{li2021invisible}
Y.~Li, Y.~Li, B.~Wu, L.~Li, R.~He, and S.~Lyu, ``Invisible backdoor attack with
  sample-specific triggers,'' in \emph{Proceedings of the IEEE/CVF
  International Conference on Computer Vision}, 2021, pp. 16\,463--16\,472.

\bibitem{liu2018fine}
K.~Liu, B.~Dolan-Gavitt, and S.~Garg, ``Fine-pruning: Defending against
  backdooring attacks on deep neural networks,'' in \emph{Research in Attacks,
  Intrusions, and Defenses: 21st International Symposium, RAID 2018, Heraklion,
  Crete, Greece, September 10-12, 2018, Proceedings 21}.\hskip 1em plus 0.5em
  minus 0.4em\relax Springer, 2018, pp. 273--294.

\bibitem{liu2020reflection}
Y.~Liu, X.~Ma, J.~Bailey, and F.~Lu, ``Reflection backdoor: A natural backdoor
  attack on deep neural networks,'' in \emph{Computer Vision--ECCV 2020: 16th
  European Conference, Glasgow, UK, August 23--28, 2020, Proceedings, Part X
  16}.\hskip 1em plus 0.5em minus 0.4em\relax Springer, 2020, pp. 182--199.

\bibitem{loshchilov2017decoupled}
I.~Loshchilov and F.~Hutter, ``Decoupled weight decay regularization,''
  \emph{arXiv preprint arXiv:1711.05101}, 2017.

\bibitem{lukas2022sok}
N.~Lukas, E.~Jiang, X.~Li, and F.~Kerschbaum, ``Sok: How robust is image
  classification deep neural network watermarking?'' in \emph{2022 IEEE
  Symposium on Security and Privacy (SP)}.\hskip 1em plus 0.5em minus
  0.4em\relax IEEE, 2022, pp. 787--804.

\bibitem{lukas2023ptw}
N.~Lukas and F.~Kerschbaum, ``Ptw: Pivotal tuning watermarking for pre-trained
  image generators,'' \emph{32nd USENIX Security Symposium (USENIX Security
  23)}, 2023.

\bibitem{madry2018towards}
\BIBentryALTinterwordspacing
A.~Madry, A.~Makelov, L.~Schmidt, D.~Tsipras, and A.~Vladu, ``Towards deep
  learning models resistant to adversarial attacks,'' in \emph{International
  Conference on Learning Representations}, 2018. [Online]. Available:
  \url{https://openreview.net/forum?id=rJzIBfZAb}
\BIBentrySTDinterwordspacing

\bibitem{may2023salient}
B.~B. May, N.~J. Tatro, P.~Kumar, and N.~Shnidman, ``Salient conditional
  diffusion for defending against backdoor attacks,'' \emph{arXiv preprint
  arXiv:2301.13862}, 2023.

\bibitem{moosavi2017universal}
S.-M. Moosavi-Dezfooli, A.~Fawzi, O.~Fawzi, and P.~Frossard, ``Universal
  adversarial perturbations,'' in \emph{Proceedings of the IEEE conference on
  computer vision and pattern recognition}, 2017, pp. 1765--1773.

\bibitem{morcos2018insights}
A.~Morcos, M.~Raghu, and S.~Bengio, ``Insights on representational similarity
  in neural networks with canonical correlation,'' \emph{Advances in Neural
  Information Processing Systems}, vol.~31, 2018.

\bibitem{nguyen2021wanet}
A.~Nguyen and A.~Tran, ``Wanet--imperceptible warping-based backdoor attack,''
  \emph{arXiv preprint arXiv:2102.10369}, 2021.

\bibitem{noppel2022disguising}
M.~Noppel, L.~Peter, and C.~Wressnegger, ``Disguising attacks with
  explanation-aware backdoors,'' in \emph{2023 IEEE Symposium on Security and
  Privacy (SP)}.\hskip 1em plus 0.5em minus 0.4em\relax IEEE Computer Society,
  2022, pp. 996--1013.

\bibitem{gpt4tech}
OpenAI, ``Gpt-4 technical report,'' \emph{arXiv preprint arXiv:2303.08774},
  2023.

\bibitem{pang:2022:eurosp}
R.~Pang, Z.~Zhang, X.~Gao, Z.~Xi, S.~Ji, P.~Cheng, and T.~Wang, ``Trojanzoo:
  Towards unified, holistic, and practical evaluation of neural backdoors,'' in
  \emph{Proceedings of IEEE European Symposium on Security and Privacy (Euro
  S\&P)}, 2022.

\bibitem{papernot2018sok}
N.~Papernot, P.~McDaniel, A.~Sinha, and M.~P. Wellman, ``Sok: Security and
  privacy in machine learning,'' in \emph{2018 IEEE European Symposium on
  Security and Privacy (EuroS\&P)}.\hskip 1em plus 0.5em minus 0.4em\relax
  IEEE, 2018, pp. 399--414.

\bibitem{qi2023revisiting}
X.~Qi, T.~Xie, Y.~Li, S.~Mahloujifar, and P.~Mittal, ``Revisiting the
  assumption of latent separability for backdoor defenses,'' in
  \emph{International Conference on Learning Representations}, 2023.

\bibitem{qiu2021deepsweep}
H.~Qiu, Y.~Zeng, S.~Guo, T.~Zhang, M.~Qiu, and B.~Thuraisingham, ``Deepsweep:
  An evaluation framework for mitigating dnn backdoor attacks using data
  augmentation,'' in \emph{Proceedings of the 2021 ACM Asia Conference on
  Computer and Communications Security}, 2021, pp. 363--377.

\bibitem{radford2021learning}
A.~Radford, J.~W. Kim, C.~Hallacy, A.~Ramesh, G.~Goh, S.~Agarwal, G.~Sastry,
  A.~Askell, P.~Mishkin, J.~Clark \emph{et~al.}, ``Learning transferable visual
  models from natural language supervision,'' in \emph{International conference
  on machine learning}.\hskip 1em plus 0.5em minus 0.4em\relax PMLR, 2021, pp.
  8748--8763.

\bibitem{ramesh2022hierarchical}
A.~Ramesh, P.~Dhariwal, A.~Nichol, C.~Chu, and M.~Chen, ``Hierarchical
  text-conditional image generation with clip latents,'' \emph{arXiv preprint
  arXiv:2204.06125}, 2022.

\bibitem{rasheed2022explainable}
K.~Rasheed, A.~Qayyum, M.~Ghaly, A.~Al-Fuqaha, A.~Razi, and J.~Qadir,
  ``Explainable, trustworthy, and ethical machine learning for healthcare: A
  survey,'' \emph{Computers in Biology and Medicine}, p. 106043, 2022.

\bibitem{roich2022pivotal}
D.~Roich, R.~Mokady, A.~H. Bermano, and D.~Cohen-Or, ``Pivotal tuning for
  latent-based editing of real images,'' \emph{ACM Transactions on Graphics
  (TOG)}, vol.~42, no.~1, pp. 1--13, 2022.

\bibitem{Sagawa2020Distributionally}
\BIBentryALTinterwordspacing
S.~Sagawa*, P.~W. Koh*, T.~B. Hashimoto, and P.~Liang, ``Distributionally
  robust neural networks,'' in \emph{International Conference on Learning
  Representations}, 2020. [Online]. Available:
  \url{https://openreview.net/forum?id=ryxGuJrFvS}
\BIBentrySTDinterwordspacing

\bibitem{salman2022does}
H.~Salman, S.~Jain, A.~Ilyas, L.~Engstrom, E.~Wong, and A.~Madry, ``When does
  bias transfer in transfer learning?'' \emph{arXiv preprint arXiv:2207.02842},
  2022.

\bibitem{sandoval2022poisons}
P.~Sandoval-Segura, V.~Singla, L.~Fowl, J.~Geiping, M.~Goldblum, D.~Jacobs, and
  T.~Goldstein, ``Poisons that are learned faster are more effective,'' in
  \emph{Proceedings of the IEEE/CVF Conference on Computer Vision and Pattern
  Recognition}, 2022, pp. 198--205.

\bibitem{schulth2022detecting}
L.~Schulth, C.~Berghoff, and M.~Neu, ``Detecting backdoor poisoning attacks on
  deep neural networks by heatmap clustering,'' \emph{arXiv preprint
  arXiv:2204.12848}, 2022.

\bibitem{schwarzschild2021just}
A.~Schwarzschild, M.~Goldblum, A.~Gupta, J.~P. Dickerson, and T.~Goldstein,
  ``Just how toxic is data poisoning? a unified benchmark for backdoor and data
  poisoning attacks,'' in \emph{International Conference on Machine
  Learning}.\hskip 1em plus 0.5em minus 0.4em\relax PMLR, 2021, pp. 9389--9398.

\bibitem{shafahi2018poison}
A.~Shafahi, W.~R. Huang, M.~Najibi, O.~Suciu, C.~Studer, T.~Dumitras, and
  T.~Goldstein, ``Poison frogs! targeted clean-label poisoning attacks on
  neural networks,'' \emph{Advances in neural information processing systems},
  vol.~31, 2018.

\bibitem{shan2022poison}
S.~Shan, A.~N. Bhagoji, H.~Zheng, and B.~Y. Zhao, ``Poison forensics: Traceback
  of data poisoning attacks in neural networks,'' in \emph{31st USENIX Security
  Symposium (USENIX Security 22)}, 2022, pp. 3575--3592.

\bibitem{shen2022sok}
J.~Shen, N.~Wang, Z.~Wan, Y.~Luo, T.~Sato, Z.~Hu, X.~Zhang, S.~Guo, Z.~Zhong,
  K.~Li \emph{et~al.}, ``Sok: On the semantic ai security in autonomous
  driving,'' \emph{arXiv preprint arXiv:2203.05314}, 2022.

\bibitem{steinhardt2017certified}
J.~Steinhardt, P.~W.~W. Koh, and P.~S. Liang, ``Certified defenses for data
  poisoning attacks,'' \emph{Advances in neural information processing
  systems}, vol.~30, 2017.

\bibitem{tang2021demon}
D.~Tang, X.~Wang, H.~Tang, and K.~Zhang, ``Demon in the variant: Statistical
  analysis of dnns for robust backdoor contamination detection.'' in
  \emph{USENIX Security Symposium}, 2021, pp. 1541--1558.

\bibitem{tao2022better}
G.~Tao, G.~Shen, Y.~Liu, S.~An, Q.~Xu, S.~Ma, P.~Li, and X.~Zhang, ``Better
  trigger inversion optimization in backdoor scanning,'' in \emph{Proceedings
  of the IEEE/CVF Conference on Computer Vision and Pattern Recognition}, 2022,
  pp. 13\,368--13\,378.

\bibitem{Tran2018SpectralSI}
B.~Tran, J.~Li, and A.~Madry, ``Spectral signatures in backdoor attacks,'' in
  \emph{Neural Information Processing Systems}, 2018.

\bibitem{tran2018spectral}
------, ``Spectral signatures in backdoor attacks,'' \emph{Advances in neural
  information processing systems}, vol.~31, 2018.

\bibitem{turner2018clean}
A.~Turner, D.~Tsipras, and A.~Madry, ``Clean-label backdoor attacks,'' 2018.

\bibitem{udeshi2022model}
S.~Udeshi, S.~Peng, G.~Woo, L.~Loh, L.~Rawshan, and S.~Chattopadhyay, ``Model
  agnostic defence against backdoor attacks in machine learning,'' \emph{IEEE
  Transactions on Reliability}, vol.~71, no.~2, pp. 880--895, 2022.

\bibitem{wang2019neural}
B.~Wang, Y.~Yao, S.~Shan, H.~Li, B.~Viswanath, H.~Zheng, and B.~Y. Zhao,
  ``Neural cleanse: Identifying and mitigating backdoor attacks in neural
  networks,'' in \emph{2019 IEEE Symposium on Security and Privacy (SP)}.\hskip
  1em plus 0.5em minus 0.4em\relax IEEE, 2019, pp. 707--723.

\bibitem{wangtraining}
Z.~Wang, H.~Ding, J.~Zhai, and S.~Ma, ``Training with more confidence:
  Mitigating injected and natural backdoors during training,'' in
  \emph{Advances in Neural Information Processing Systems}.

\bibitem{wang2022training}
------, ``Training with more confidence: Mitigating injected and natural
  backdoors during training,'' \emph{Advances in Neural Information Processing
  Systems}, vol.~35, pp. 36\,396--36\,410, 2022.

\bibitem{Weber2020RABPR}
M.~Weber, X.~Xu, B.~Karlas, C.~Zhang, and B.~Li, ``Rab: Provable robustness
  against backdoor attacks,'' \emph{ArXiv}, vol. abs/2003.08904, 2020.

\bibitem{wenger2022finding}
E.~Wenger, R.~Bhattacharjee, A.~N. Bhagoji, J.~Passananti, E.~Andere, H.~Zheng,
  and B.~Zhao, ``Finding naturally occurring physical backdoors in image
  datasets,'' \emph{Advances in Neural Information Processing Systems},
  vol.~35, pp. 22\,103--22\,116, 2022.

\bibitem{wu2022backdoorbench}
B.~Wu, H.~Chen, M.~Zhang, Z.~Zhu, S.~Wei, D.~Yuan, C.~Shen, and H.~Zha,
  ``Backdoorbench: A comprehensive benchmark of backdoor learning,''
  \emph{arXiv preprint arXiv:2206.12654}, 2022.

\bibitem{wu2021adversarial}
D.~Wu and Y.~Wang, ``Adversarial neuron pruning purifies backdoored deep
  models,'' \emph{Advances in Neural Information Processing Systems}, vol.~34,
  pp. 16\,913--16\,925, 2021.

\bibitem{xiang2022patchcleanser}
C.~Xiang, S.~Mahloujifar, and P.~Mittal, ``$\{$PatchCleanser$\}$: Certifiably
  robust defense against adversarial patches for any image classifier,'' in
  \emph{31st USENIX Security Symposium (USENIX Security 22)}, 2022, pp.
  2065--2082.

\bibitem{yao2019latent}
Y.~Yao, H.~Li, H.~Zheng, and B.~Y. Zhao, ``Latent backdoor attacks on deep
  neural networks,'' in \emph{Proceedings of the 2019 ACM SIGSAC Conference on
  Computer and Communications Security}, 2019, pp. 2041--2055.

\bibitem{zhang2018unreasonable}
R.~Zhang, P.~Isola, A.~A. Efros, E.~Shechtman, and O.~Wang, ``The unreasonable
  effectiveness of deep features as a perceptual metric,'' in \emph{Proceedings
  of the IEEE conference on computer vision and pattern recognition}, 2018, pp.
  586--595.

\bibitem{zhang2022bagflip}
Y.~Zhang, A.~Albarghouthi, and L.~D'Antoni, ``Bagflip: A certified defense
  against data poisoning,'' \emph{arXiv preprint arXiv:2205.13634}, 2022.

\bibitem{zhu2019transferable}
C.~Zhu, W.~R. Huang, H.~Li, G.~Taylor, C.~Studer, and T.~Goldstein,
  ``Transferable clean-label poisoning attacks on deep neural nets,'' in
  \emph{International Conference on Machine Learning}.\hskip 1em plus 0.5em
  minus 0.4em\relax PMLR, 2019, pp. 7614--7623.

\end{thebibliography}

\appendix

\section{Appendix}

% ++++++++++++++++++++++++++++++++++++++++++++++++++++++++++++++++++++
\subsection{Model Architectures}
\label{appendix:model-architecture}
% ++++++++++++++++++++++++++++++++++++++++++++++++++++++++++++++++++++
We use a ResNet-18~\cite{he2016deep} architecture with a clean data accuracy of 95\% on CIFAR-10.
For ImageNet, we use two models: a pre-trained ResNet-18 model with a clean data accuracy of 70\% and OpenAI's pre-trained \texttt{clip-vit-base-patch32}\footnote{\url{https://huggingface.co/openai/clip-vit-base-patch32}} model checkpoint.

% ------------------------------------
\subsection{Summary of the Surveyed Attacks}
\label{appendix:surveyed_attack_summary}
% ------------------------------------

\begin{figure*}
    \centering
    \includegraphics[width=1.\linewidth]{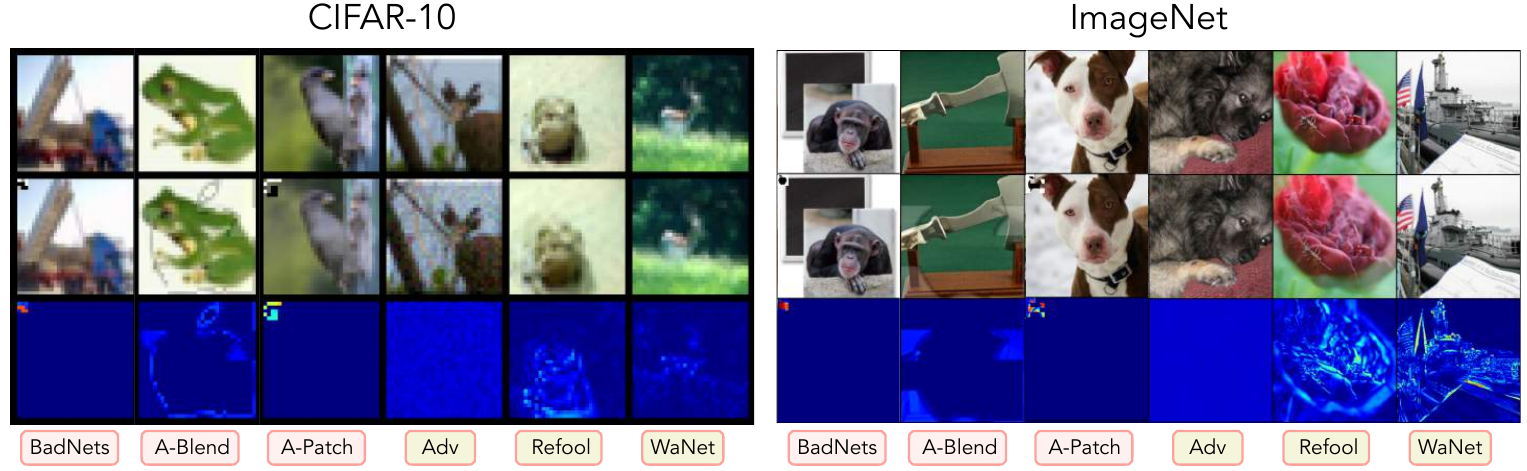}
    \caption{Three rows showing clean and poisoned samples in the top two rows and a heatmap of their differences on the bottom row. We highlight \poisonlabel{poison-label} and \cleanlabel{clean-label} attacks and summarize the hyper-parameters in \Cref{appendix:surveyed_attack_summary}.}
    \label{fig:triggers}
\end{figure*}
We re-implement all surveyed data poisoning attacks from scratch and distinguish between \cleanlabel{clean-label} and \poisonlabel{poison-label} attacks.
This section first describes all surveyed data poisoning attacks and then provides details on the hyper-parameters used for studying the data efficiency of our defense in \Cref{tab:data-efficiency-table}. 

\poisonlabel{BadNets}~\cite{gu2017badnets}: This attack uses a visible, patch-shaped trigger pattern that is stamped on the image. 
In our experiments, we downscale the image of a black apple on a white background to the size of the trigger and stamp this trigger on the top left corner of the image. 
Throughout our paper, we use a trigger pattern with size $\lfloor \nicefrac{3}{32} \cdot w \rfloor$ pixels relative to the image size $w$, covering less than $1\%$ of the image's total area.  

\poisonlabel{Adaptive Patch}~\cite{qi2023revisiting}: This attack is an extension of \poisonlabel{BadNets} that (i) places a grid on top of the trigger with $s\times s$ squares and (ii) randomly occludes squares for each poisoned image. 
We use a grid size of $s=4$ with an occlusion probability per square of $50\%$. 
The authors distinguish between \emph{payload} and \emph{regularization} samples. 
A payload sample will be assigned the target label, whereas a regularization sample will receive the ground-truth label (despite containing the trigger). 
The authors argue that this is done to ensure that the poisoned and target samples are not separable in the poisoned model's latent space. 
They use a \emph{conservatism ratio} to control the ratio between payload and regularization samples. 
We use a conservatism ratio of $0.5$. 

\poisonlabel{Adaptive Blend}~\cite{qi2023revisiting}: This is a variation of the \poisonlabel{Adaptive Patch} attack, but instead of placing a patch on the top left corner of the image, the trigger is equivalent to the size of the image, but it is superimposed by blending it using an opacity $\alpha\in [0, 1]$. 
The authors use the same trick as before and partially occlude the trigger.
We use $s=4$, an occlusion rate of $50\%$, and an opacity $\alpha=0.3$.
This attack uses the same trick as the \poisonlabel{Adaptive Patch} attack to prevent latent separability. 
We also use a conservatism ratio of $0.5$ for this attack.

\cleanlabel{AdvClean}~\cite{turner2018clean}: This attack assumes access to a similar \emph{base} model trained on the same task. 
For both CIFAR-10 and ImageNet, we use a base model provided by the torchvision package\footnote{\url{https://pytorch.org/vision/stable/models.html}} that has been trained on ImageNet and has a ResNet-18 architecture (the same architecture as the attacked model). 
To create a trigger, the attacker generates targeted adversarial examples using the base model against some class (we always use the target class 421 - \textit{banister}) using the Projected Gradient Descent (PGD) attack ~\cite{madry2018towards}. 
We iterate PGD for four steps using $\epsilon=\nicefrac{8}{255}$. 

\cleanlabel{Refool}~\cite{liu2020reflection}: This attack uses naturally occurring phenomenons as the trigger pattern, such as reflections. 
We attack the target class \textit{swing} and create the ghosting effect as implemented by the \texttt{trojanzoo} library~\cite{pang:2022:eurosp}. 
Our experiments show that using small values for the ghosting intensity does not lead to effective attacks. 
For this reason, we use larger values to create the ghosting effect, specifically $\alpha=1$, which creates blurry artifacts as illustrated in \Cref{fig:triggers}.
We verified that a non-poisoned model still correctly predicts a majority of these images, but note that our chosen hyper-parameters could make the generated samples more easily detectable during dataset sanitation.  
The authors do not present results on ImageNet.

\cleanlabel{WaNet}~\cite{nguyen2021wanet}: This attack uses a warping effect on the image, which preserves the image's perceptual quality (as opposed to \cleanlabel{Refool}). 
We re-use the original implementation by the authors\footnote{\url{https://github.com/VinAIResearch/Warping-based_Backdoor_Attack-release}} but again have to use stronger perturbations as those proposed by the authors. 
We use about three times as much noise as the authors use in their paper.  
The authors do not show results on ImageNet.

\subsubsection{Training Poisoned Models.}
We use the same hyper-parameters on each dataset to train the models on poisoned data for all attacks, mirroring an attack scenario in practice. 
We always train on 100\% of the training data and poison \emph{without replacement}, i.e., a clean-label attack does not modify the class distribution in the training dataset (i.e., the number of samples per class remains constant).
The hyperparameters for the trainer are listed below in \Cref{tab:training_parameters}.  
As outlined in the paper, we only fine-tune models on ImageNet instead of training from scratch on CIFAR-10. 

\begin{table}[ht]
\centering
\caption{Training Parameters for CIFAR-10 and ImageNet}
\label{tab:training_parameters}
\begin{tabular}{lcc}
\hline
Parameter                     & CIFAR-10 & ImageNet \\
\hline
Epochs                        & 120      & 10        \\
Momentum                      & 0.9      & 0.9      \\
Learning rate (lr)            & 0.1      & 0.0001   \\
Weight decay                  & 0.0005   & 0.0001   \\
Cosine annealing scheduler    & True     & False     \\
T\_max                        & 120      & -       \\
Optimizer                     & SGD      & SGD      \\
From Scratch                  & True     & False    \\
Batch size                    & 128      & 128    \\
\hline
\end{tabular}
\end{table}

% +++++++++++++++++++++++++++++++++++++
\subsection{Summary of the Surveyed Defenses}
\label{appendix:defense-description}
% +++++++++++++++++++++++++++++++++++++
\textbf{Weight Decay}~\cite{loshchilov2017decoupled}: We implement fine-tuning with weight decay, which is a well-known regularization technique for deep neural networks that adds a loss-term scaled by $\lambda_w$ to each weight proportional to its Euclidean norm.

\textbf{Fine-Pruning}~\cite{liu2018fine}: Fine-Pruning consists of three steps: (1) Measure the absolute activation of every convolutional layer on clean data, (2) select the $\rho\in [0,1]$ channels with the lowest mean absolute activations and perform a channel-wise resetting of all parameters (we reset these weights to zero).
Finally, (3) fine-tune the model on clean data to regain the test accuracy lost to pruning.

\textbf{Neural Cleanse}~\cite{wang2019neural}: Neural Cleanse can be instantiated as a backdoor detection or model repair defense. 
It consists of three steps that are encoded by \Cref{alg:neural-cleanse}: (1) reverse-engineer a trigger pattern for each class, (2) compute the L1-norm for each trigger and return the lowest norm as an indication of whether the model has been backdoored and (3) if the objective is to repair the model, fine-tune the model on a mix of clean and poisoned data (using the reverse-engineered trigger). 

\begin{figure}
    \centering
    \includegraphics[width=1.\linewidth]{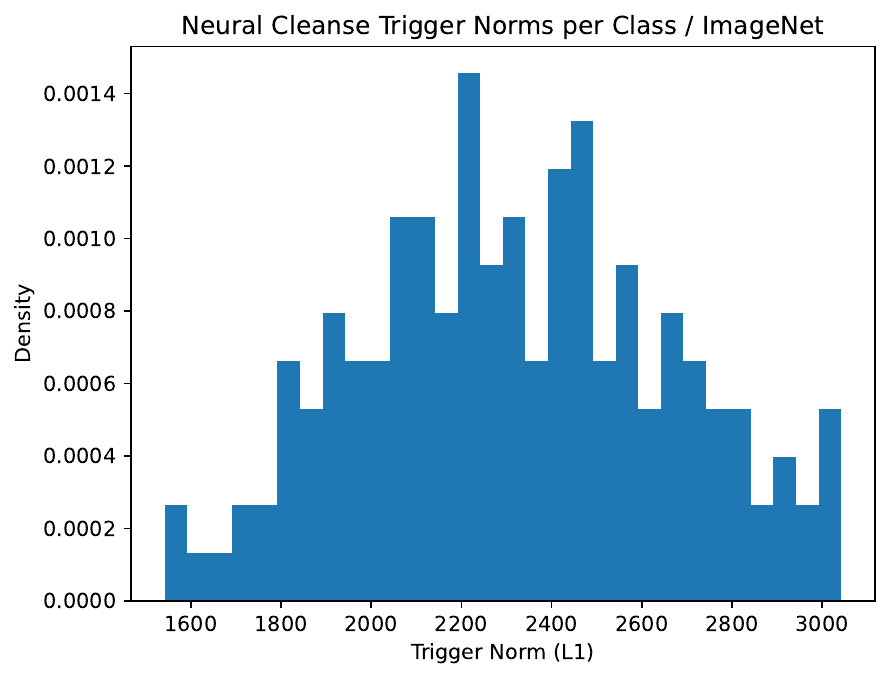}
    \includegraphics[width=1.\linewidth]{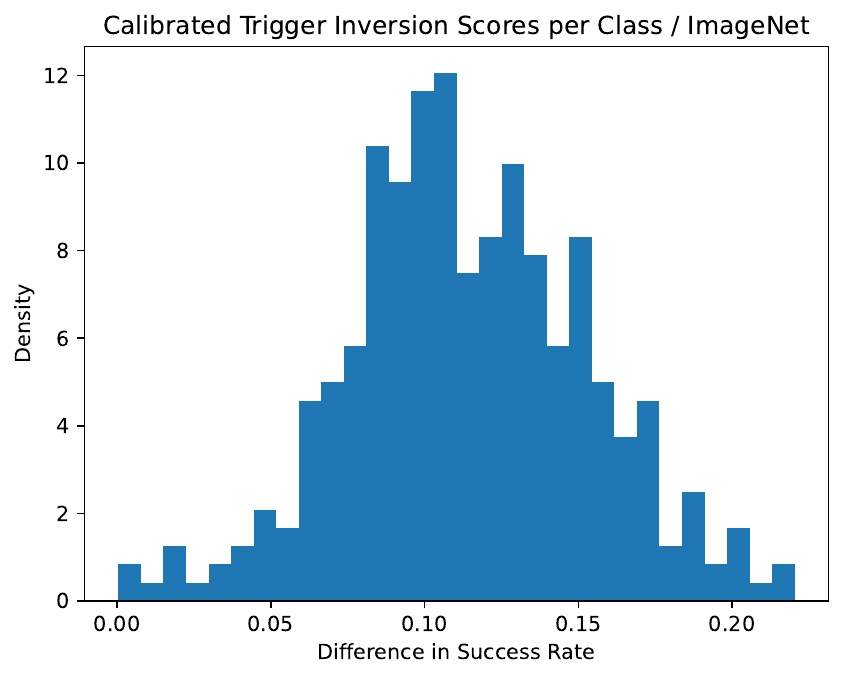}
    \caption{The anomaly scores on ImageNet of reverse-engineered trigger for Neural Cleanse~\cite{li2021neural} (top) and our Calibrated Trigger Inversion method (bottom), measured by the difference in success rate for the model before and after repair using our Pivotal Tuning-based defense with $r=5\%$. We iterate over all classes and store the anomaly score for the reversed trigger for each class. This statistic is useful for detecting anomalies (i.e., backdoored classes). }
    \label{fig:neural-cleanse-trigger}
\end{figure}

We calibrate Neural Cleanse by computing the trigger norm for each class, depicted in \Cref{fig:neural-cleanse-trigger}. 
As opposed to the authors, who only provide constant bounds for detection, which are dataset-specific and cannot be applied to our use case, whenever we do not use ROC AUC, we apply a statistical test to estimate the probability that this measurement has not been sampled from this normal distribution. 

\textbf{Neural Attention Distillation (NAD)}~\cite{li2021nad}: NAD consists of two steps: (1) Clone the poisoned model, which we refer to as a teacher model, and fine-tune it on the trusted data for a limited number of steps. 
(2) Fine-tune the poisoned student model using the attention loss presented in the author's paper on each neural network layer to align the latent spaces between the student and teacher models on the trusted data.

\subsection{Adaptive Attacks}
\Cref{alg:tsb-attack} encodes our adaptive Trigger-Scattering Backdoor (\poisonlabel{TSB}) attack. 
The data poisoning attack receives an image $x$, a target label $y_{\text{target}}\in \mathcal{Y}$, and $k$ perceptually distinct, secret trigger patterns as input and returns a poisoned image. 
We distinguish between two cases: The attack adds a single trigger to poisoned samples injected before the model's training (lines 2-4) and multiple triggers on samples used to exploit the backdoor during inference (lines 5-8).  

\begin{algorithm}
\caption{Neural Cleanse~\cite{wang2019neural}}
\begin{algorithmic}[1]
\Procedure{Neural-Cleanse}{$\theta, D_{\text{trust}}, Y_{\text{trust}}, N_1, N_2, \alpha$}  
    \State $S \gets \{\}$ \Comment{Used to store reverse-engineered triggers}
    \For{$y_{\text{target}} \gets 1 \textbf{ to } |\mathcal{Y}|$} 
    \State $T^* \gets \text{Random Initialization}$ 
        \For{$1 \textbf{ to } N_1$}
            \State $x \sim D_{\text{trust}}$ 
            \State $y_{\text{pred}} \gets M(E(x \oplus T^*;\theta);\theta)$
            \State $g_{T^*} \gets \nabla_{T^*} \text{CE}(y_{\text{pred}}, y_{\text{target}})$ \Comment{Classify}
            \State $T^* \gets T^* - \alpha \cdot \text{Adam}(T^*, g_{T^*})$ \Comment{Update trigger}
        \EndFor
        \State $S \gets S \cup \{T^*\}$ 
    \EndFor
    \State $i \gets \operatorname{argmin} \{\left\lVert S_i \right\rVert_1|S_i \in S\}$ \Comment{Find lowest norm}
    \BeginBox[draw=black,dashed]
    \If{\text{isDetection}} \Comment{Detection case}
        \Return $\left\lVert S_i \right\rVert_1$
    \EndIf
    \EndBox
    
    \For{$j\gets 1 \textbf{ to } N_2$} \Comment{Model repair case}
            \State $x, y_{\text{target}} \sim D_{\text{trust}} \times Y_{\text{trust}}$ 
            \State $\hat{x} \gets (x \oplus S_i) \text{ if j is even else } x$
            \State $y_{\text{pred}} \gets M(E(\hat{x};\theta);\theta)$ \Comment{Classify}
            \State $g_{\theta} \gets \nabla_{\theta} \text{CE}(y_{\text{pred}}, y_{\text{target}})$ 
            \State $\theta \gets \theta - \alpha \cdot \text{Adam}(\theta, g_{\theta})$ \Comment{Update model}
        \EndFor
    \Return $\theta$
\EndProcedure
\end{algorithmic}
\label{alg:neural-cleanse}
\end{algorithm}

\begin{algorithm}
\caption{Trigger-Scattering Backdoor (TSB) / Adaptive Attack}
\begin{algorithmic}[1]
\Procedure{$\mathcal{A}_{\poisonlabel{TSB}}$}{$x, y_\text{target}, T_1,..,T_k,$} 
    \BeginBox[draw=black,dashed]
    \If{\text{injectSamples}} \Comment{Data poisoning / Before training}
        \State $i \sim \{1,.., |D_\text{clean}|\}$
        \Return $x \oplus T_i, y_\text{target} $\Comment{Add trigger at a random location}
    \EndIf
    \EndBox
    \State $\Tilde{x} \gets x$ 
    \For{$j \gets 1 \textbf{ to } k$}
        \State $\Tilde{x} \gets \Tilde{x} \oplus T_j$ \Comment{Add all triggers to exploit the backdoor}
    \EndFor
    \Return $\Tilde{x}, y_\text{target}$
\EndProcedure
\end{algorithmic}
\label{alg:tsb-attack}
\end{algorithm}

% +++++++++++++++++++++++++++++++++++++
\subsection{Defense Hyper-Parameter Search}
\label{appendix:used-defense-parameters}
% +++++++++++++++++++++++++++++++++++++
\begin{table}[ht]
\centering
\caption{Results of hyper-parameter search for all surveyed defenses on CIFAR-10 and ImageNet using $r=1\%$.}
\begin{tabular}{@{}lcc@{}}
\toprule
Parameter & CIFAR-10 & ImageNet \\ \midrule
\multicolumn{3}{c}{Pivotal Tuning} \\
n\_steps / $N$ & 2,000 & 10,000 \\
opt & sgd & sgd \\
lr / $\alpha$ & 0.01 & 5e-4 \\
slol\_lambda / $\lambda_S$ & 0.05 & 1e-5 \\
param\_lambda / $\lambda_R$ & 20,000 & 150,000 \\
weight\_decay & 0 & 0 \\
batch\_size & 128 & 128 \\
\midrule
\multicolumn{3}{c}{Neural Attention Distillation} \\
n\_steps / $N$ & 2,000 & 8,000 \\
opt & sgd & sgd \\
lr / $\alpha$ & 0.01 & 5e-4 \\
teacher\_steps & 1,000 & 1,000 \\
power / $p$ & 2 & 2 \\
at\_lambda / $\lambda_{\text{at}}$ & 1,000 & 1,000 \\
weight\_decay & 0 & 0 \\
batch\_size & 128 & 128 \\
\midrule
\multicolumn{3}{c}{Neural Cleanse} \\
n\_steps / $N$ & 1,000 & 3,000 \\
opt & sgd & sgd \\
lr / $\alpha$ & 0.001 & 5e-4 \\
steps\_per\_class / $N_1$       & 100         & 200                  \\
        norm\_lambda / $\lambda_N$     &   2e-2        & 1e-5              \\
        weight\_decay   & 0                & 0           \\ 
        batch\_size   & 128                & 128           \\ 
\midrule
\multicolumn{3}{c}{Fine-Pruning} \\
n\_steps / $N$     & 1,000           & 5,000                   \\
        opt            & sgd              & sgd                         \\
        lr / $\alpha$      & 0.001          & 5e-4               \\
        prune\_rate / $\rho$    & $96\%$         & $73\%$                  \\
        sampled\_batches     & 10         & 10                 \\
        weight\_decay   & 0                & 0           \\ 
        batch\_size   & 128                & 128           \\ 
\midrule
\multicolumn{3}{c}{Weight-Decay} \\
n\_steps / $N$     & 1,000           & 5,000                   \\
        opt            & sgd              & sgd                         \\
        lr / $\alpha$      & 0.001          & 5e-4               \\
        weight\_decay   & 0.01                & 0.001           \\ 
        batch\_size   & 128                & 128           \\ 
\bottomrule
\end{tabular}
\label{tab:combined-parameters}
\end{table}
This section summarizes the parameters and our notation used in the hyper-parameter grid search for the five surveyed post-training defenses described in \Cref{appendix:defense-description}. 
As described in the main part of the paper, we find optimal hyper-parameters by poisoning the model with a BadNets~\cite{gu2017badnets} backdoor through fine-tuning with the trusted data. 
This mirrors a realistic scenario, where the defender has access to the pre-trained model weights and limited trusted data, meaning they are capable of poisoning their own model.

We search the following space of hyper-parameter for each defense.
For all defenses, we experiment with an SGD and Adam optimizer, a learning rate of $\alpha \in [0.1,..,1\text{e-3}]$ on CIFAR-10 and $\alpha \in [1\text{e-3},..,1\text{e-6}]$ on ImageNet.
For each defense, we ablate over the following hyper-parameters.
For brevity, we only provide the ranges that we checked instead of providing every combination of each parameter. 
We evaluate over 20 configurations for each defense on each dataset. 
\begin{itemize}
    \item \textbf{Pivotal Tuning} (Ours): On CIFAR-10, we search over $\lambda_S \in [0.1,..,0.001]$ and on ImageNet we use $\lambda_S \in [1\text{e-2},..,1\text{e-6}]$.
    For the parameter regularization $\lambda_R$, we search over $\lambda_R \in [10^6,..,2\cdot 10^7]$ for both datasets. 
    \item \textbf{Weight Decay}: We search over the strength of the weight decay $\lambda_{w}\in \{0.01,..,1\text{e-5}\}$.
    \item \textbf{NAD}~\cite{li2021nad}: 
    We search over the steps used to fine-tune the teacher $N_0 \in \{1000,.., 2000\}$, the exponent $p\in \{2,..,4\}$ and the strength for the feature alignment $\lambda_{\text{at}}\in \{100,..,10\,000\}$.
    \item \textbf{Neural Cleanse}~\cite{wang2019neural}: We search over the penalty for the trigger norm $\lambda_N \in \{2\text{e-2},..,1\text{e-6}\}$.
    \item \textbf{Fine-Pruning}~\cite{liu2018fine}: We apply channel-wise pruning with a ratio $\rho \in \{0.5, .., 0.99\}$.
\end{itemize}

\end{document}